\documentclass[final,twocolumn]{raa_twocolumn}           
\usepackage{graphicx,times}
\usepackage{natbib}
\usepackage{amssymb,amsmath}
\usepackage{multirow}
\usepackage[version=4]{mhchem}
\usepackage{txfonts}
\usepackage{mathtools}
\usepackage{adjustbox}

\bibpunct{(}{)}{;}{a}{}{,}

\usepackage[colorlinks=true, allcolors=blue]{hyperref}

\usepackage{ulem}

\usepackage{CJK}

\usepackage{xcolor}
\definecolor{addnewseccolor}{rgb}{0.5, 0.3, 0.2}
\begin{document}
\begin{CJK*}{UTF8}{gbsn}

\title{Organic Acid Chemistry in ISM: Detection of Formic Acid and its Prebiotic Chemistry in Hot Core G358.93--0.03 MM1}

\volnopage{ {\bf 2024} Vol.\ {\bf X} No. {\bf XX}, 000--000}
  \setcounter{page}{1}

\author{
Arijit Manna\thanks{amanna.astro@gmail.com} \inst{1}, Sabyasachi Pal \inst{1}, Sekhar Sinha \inst{2}, Sushanta Kumar Mondal \inst{2}}

\institute{ 
$^{1}$Department of Physics and Astronomy, Midnapore City College, Kuturia, Bhadutala, Paschim Medinipur 721129, India \\
$^{2}$Department of Physics, Sidho-Kanho-Birsha University, Ranchi Road, Purulia 723104, India\\
\vs \no
   {\small Received 20XX Month Day; accepted 2025 Jan 29}
}

\abstract{
In the interstellar medium, formic acid (HCOOH) plays a significant role in the synthesis of the simplest amino acid, glycine (NH$_{2}$CH$_{2}$COOH). The presence of HCOOH suggests that oxygen-bearing molecules may be directly involved in the chemical and physical evolution of star formation regions, particularly in hot molecular cores. This paper presents the first detection of the rotational emission lines of the $trans$-conformer of HCOOH toward the hot molecular core G358.93--0.03 MM1, located in the massive star formation region G358.93--0.03. This study employed high-resolution observations from the Atacama Large Millimeter/submillimeter Array (ALMA) in Band 7. The column density and excitation temperature of $t$-HCOOH are determined as $(8.13\pm0.72)\times10^{15}$ cm$^{-2}$ and $120\pm15$ K, respectively. The fractional abundance of $t$-HCOOH relative to H$_{2}$ is $(2.62\pm 0.29)\times 10^{-9}$. The column density ratios of $t$-HCOOH/\ce{CH3OH} and $t$-HCOOH/\ce{H2CO} are $(1.56 \pm 0.12)\times 10^{-2}$ and $(1.16 \pm 0.12)$, respectively. We computed a three-phase warm-up chemical model of HCOOH using the gas-grain chemical code UCLCHEM. We found that the observed and modelled abundances of HCOOH are almost identical, within a factor of 0.89. Based on chemical modelling, we showed that HCOOH may be formed through the reaction between HCO and OH on the grain surface, which is further released in the gas-phase.
\keywords{ISM: individual objects (G358.93--0.03) -- ISM: abundances -- ISM: kinematics and dynamics -- stars: formation -- astrochemistry}}

\authorrunning{Manna et al.}            
\titlerunning{Formic acid in G358.93--0.03}  
   \maketitle

%
\section{Introduction}           
\label{sect:intro}
The number of discoveries of interstellar complex organic molecules (COMs) has increased significantly over the last ten years. In the interstellar medium (ISM) or circumstellar shells, more than 320\footnote{\url{https://cdms.astro.uni-koeln.de/classic/molecules}} molecular compounds have been found, including amines (R--NH$_{2}$), carboxamides (R--C(=O)--NR$_{2}$), alcohols (C$_{n}$H$_{2n+1}$OH), acids (--COOH), and aldehydes (R--CH=O) \citep{gu22}. Various types of COMs in ISM represent the essential functional groups needed to initiate the synthesis of ribonucleic acid (RNA) and prebiotic compounds \citep{redo15, cha20, her22}. Among the detected COMs, various organic acids from different sources have been observed in the gas-phase at low abundances relative to water (H$_{2}$O) \citep{gu22}. Previous chemical models have indicated that acids play a vital role in the formation of more complex biomolecules, such as amino acids (R--CH(NH$_{2}$)--COOH), in star formation regions, molecular clouds, and planetary atmospheres \citep{eh00, gar06, ko23}.

\begin{figure}
\centering
\includegraphics[width=0.5\textwidth]{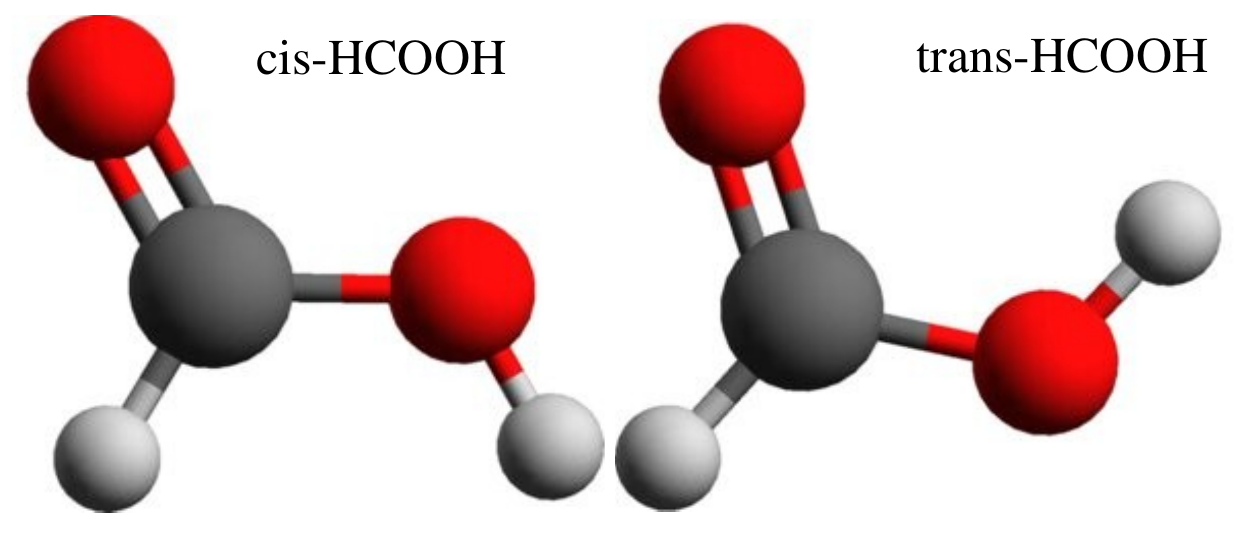}
\caption{Molecular structures of $cis$-HCOOH and $trans$-HCOOH. Carbon (C), hydrogen (H), and oxygen (O) atoms are represented by grey, white, and red spheres, respectively.}
\label{fig:moldia}
\end{figure}

Among the different types of COMs, formic acid (HCOOH) is one of the most interesting molecules because it is found in ant bites, stinging nettle plants, and bee stings. HCOOH is structurally similar to glycine (NH$_{2}$CH$_{2}$COOH), acetic acid (CH$_{3}$COOH), and methyl formate (CH$_{3}$OCHO). The recent quantum chemical model of \cite{pin25} showed that HCOOH produces \ce{NH2CH2COOH} in the gas-phase via the following chemical reaction.\\\\
HCOOH + \ce{NH2CH} $\longrightarrow$ \ce{NH2CH2COOH}~~~~~~~~~~(1)\\\\
In Reaction 1, \ce{NH2CH} is one of the isomers of methyleneimine (\ce{CH2NH}), which is formed via the decomposition of \ce{CH2NH} under UV radiation \citep{bou23}. Reaction 1 is one of the ideal pathways to the formation of \ce{NH2CH2COOH} because HCOOH and \ce{CH2NH} are both found in ISM \citep{peng19, suz23, ma24}. Although HCOOH is related to these biologically significant species, it has not received much observational attention. According to the usual astronomical classification, HCOOH is not completely a COM because it contains fewer than six atoms \citep{bel13, jo20, her22}. Since HCOOH has an impact on the atmospheric environment, the human body, and the creation of interstellar COMs, it is a particularly significant organic acid \citep{ca14}. Quantum chemical studies indicate that HCOOH has two conformers, such as $trans$ and $cis$, based on the orientation of the single hydrogen bond \citep{ho76}. The three-dimensional molecular structures of the $cis$ and $trans$ conformers of HCOOH are shown in Figure~\ref{fig:moldia}, adapted from \citet{ja18} and reproduced using the molecular modelling software MolView \citep{ber15}. Among both conformers, the $trans$ conformer of HCOOH (hereafter $t$-HCOOH) is the most stable, and this conformer was also first found in Sgr B2 with an estimated column density of $\sim 3\times10^{15}$ cm$^{-2}$ \citep{zu71}. The energy of the ground-vibrational state of $cis$-HCOOH is $1365\pm30$ cm$^{-1}$, which is more than that of the $trans$ conformer \citep{ho76}. The internal rotation, or the conversion from $trans$ to $cis$, has an energy barrier of about 4827 cm$^{-1}$ ($\sim$ 7000 K) \citep{ho76}.  In ISM, HCOOH was first detected in radio frequencies towards Sgr B2 with an estimated column density of $\sim3\times10^{15}$ cm$^{-2}$ \citep{zu71, wi75}. The emission lines of HCOOH are also found towards comets \citep{biv14}, star-forming regions \citep{ik01, riv17, peng19}, hot molecular cores \citep{li01, liu02, gar22}, hot corinos \citep{caz03, jo18, os18}, molecular clouds \citep{zu71, re06, rod21}, and cold dark clouds \citep{ir90, gar23}.The sulphur (S)-bearing counterpart molecules of HCOOH are dithioformic acid (HC(S)SH) and thioformic acid (HC(O)SH), and both molecules are also detected in ISM \citep{gar22, manna23a, man24a}. 

\begin{figure*}
\centering
\includegraphics[width=1.03\textwidth]{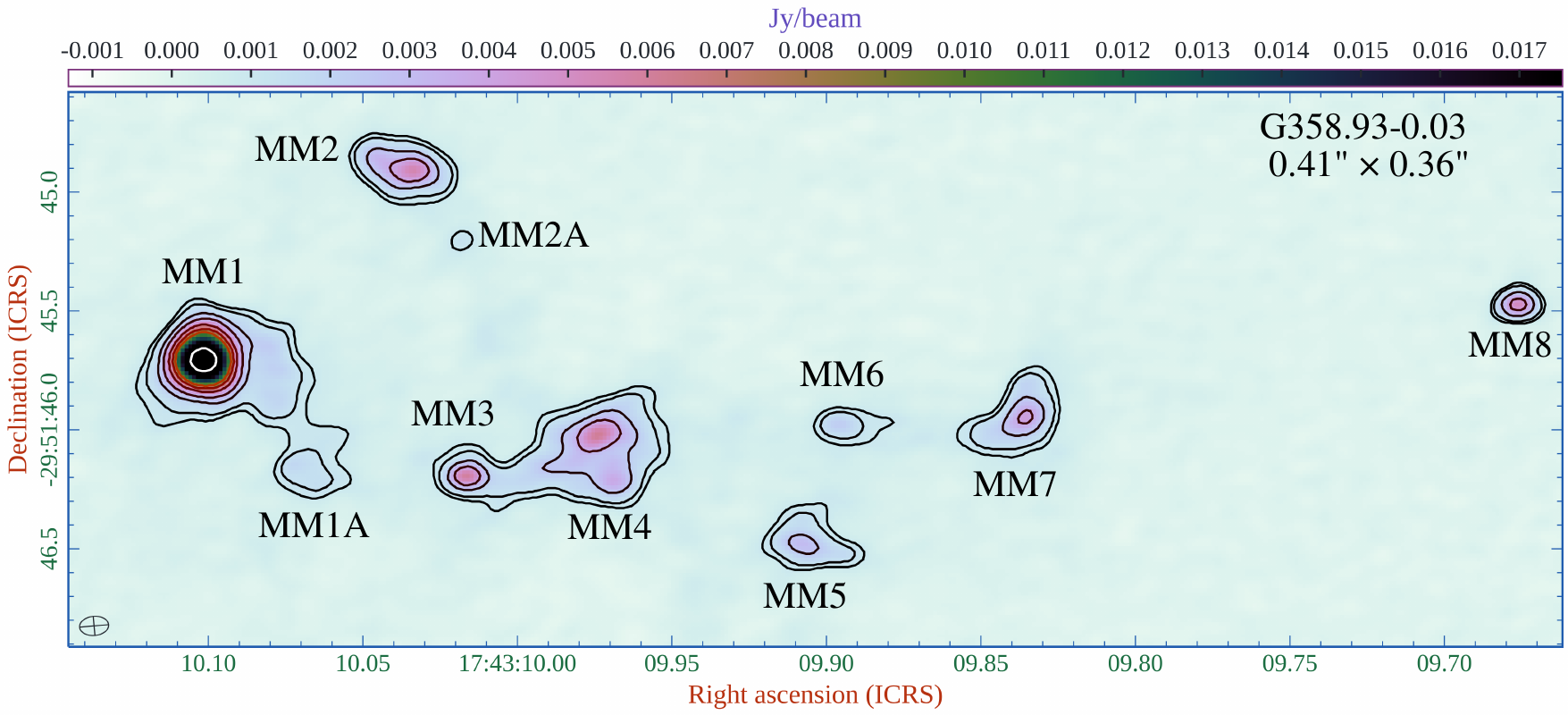}
\caption{The dust continuum emission image of the massive star formation region G358.93--0.03 at a frequency of 291.31 GHz. The synthesized beam size of the continuum image is 0.41$^{\prime\prime}$$\times$0.36$^{\prime\prime}$. The contour levels increase by a factor of $\surd$2 from the starting point of 2.5$\sigma$.}
\label{fig:dustcontinuum}
\end{figure*}

\begin{table*}
\centering
\caption{Dust continuum properties of the detected cores toward G358.93--0.03 at 291.31 GHz.}
\begin{adjustbox}{width=0.99\textwidth}
\begin{tabular}{ccccccccccccccccc}
\hline 
Source&R.A.&Decl.&Integrated flux & Peak flux &RMS&Deconvolved source size & Remark\\
	&    &     & (mJy)           &  (mJy beam$^{-1}$)&(mJy)&($^{\prime\prime}$$\times$$^{\prime\prime}$)&\\
\hline
G358.93--0.03 MM1&17:43:10.1015 &--29:51:45.7057&72.32$\pm$1.82&34.61$\pm$0.63&0.58&0.48$\times$0.39&Resolved \\
			
~~~G358.93--0.03 MM1A&17:43:10.0671&--29:51:46.4511&19.62$\pm$3.26&5.11$\pm$0.16&0.23&0.39$\times$0.34&Not resolved\\
			
G358.93--0.03 MM2&17:43:10.0357&--29:51:44.9019&22.62$\pm$2.15&7.68$\pm$0.62&0.27&0.43$\times$0.38&Resolved\\
			
~~~G358.93--0.03 MM2A&17:43:10.0209&--29:51:45.1577&16.42$\pm$1.49&3.47$\pm$0.14&0.28&0.42$\times$0.38&Resolved\\
			
G358.93--0.03 MM3&17:43:10.0145&--29:51:46.1933&62.68$\pm$2.56&19.62$\pm$2.63&0.58&0.45$\times$0.39&Resolved\\
			
G358.93--0.03 MM4&17:43:09.9738&--29:51:46.0707&50.75$\pm$2.56&12.62$\pm$0.89&0.65&0.43$\times$0.37&Resolved\\
			
G358.93--0.03 MM5&17:43:09.9063&--29.51.46.4814&22.62$\pm$1.35&2.45$\pm$0.16&0.23&0.44$\times$0.38&Resolved\\
			
G358.93--0.03 MM6&17:43:09.8962&--29:51:45.9802&12.56$\pm$0.98&3.16$\pm$0.16&0.52&0.43$\times$0.37&Resolved\\
			
G358.93--0.03 MM7&17:43:09.8365&--29:51:45.9498&11.26$\pm$0.85&2.43$\pm$0.15&0.62&0.44$\times$0.38&Resolved\\
			
G358.93--0.03 MM8&17:43:09.6761&--29:51:45.4688&9.56$\pm$0.62&1.48$\pm$0.21&0.36&0.40$\times$0.34&Not resolved\\
\hline 
\end{tabular}	
\end{adjustbox}
\label{tab:cont}
\end{table*}

\begin{figure*}
\centering
\includegraphics[width=1.03\textwidth]{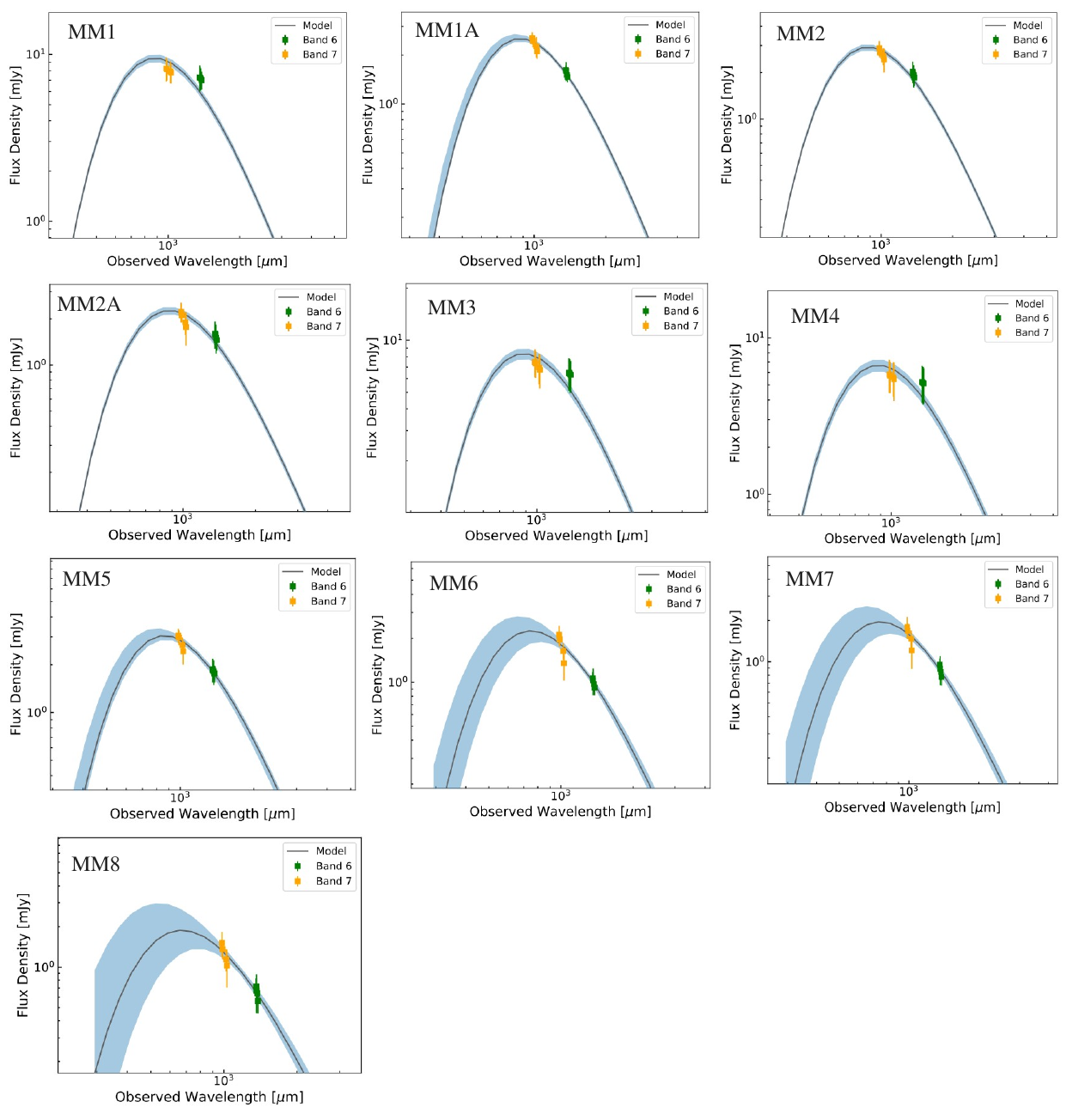}
\caption{SEDs of dust continuum cores from ALMA Band 6 and 7 observations. In the SED plots, orange and green data points represent the observed flux densities of the detected cores across different ALMA bands, with error bars indicating the measurement uncertainties. The black curve corresponds to the best-fit SED obtained using the radiative transfer model by \citet{rob07}, while the blue-shaded region illustrates the 1$\sigma$ uncertainty range of the fit.}
\label{fig:SED}
\end{figure*}

\begin{table*}
\centering
\caption{Physical properties of dust continuum sources towards G358.93--0.03.}
\begin{adjustbox}{width=1.0\textwidth} 
\begin{tabular}{ccccccccccccccccc}
\hline 
Source&Mass&Luminosity&Dust temperature ($T_{d}$) &Dust emissivity index&Spectral index$^{\dagger}$&Density$^{\ddagger}$\\
      &(\textup{M}$_{\odot}$)&(\textup{L}$_{\odot}$)&(K)&($\beta$)&($\alpha$)&(cm$^{-3}$)\\
\hline
G358.93--0.03 MM1&25.75$\pm$6.36&(7.82$\pm$1.23)$\times$10$^{4}$&175$\pm$12&1.72$\pm$0.16&3.72&1.08$\times$10$^{8}$\\
~~~G358.93--0.03 MM1A&11.26$\pm$3.67&(4.42$\pm$1.65)$\times$10$^{3}$&35$\pm$4&1.56$\pm$0.12&3.56&2.36$\times$10$^{6}$\\
G358.93--0.03 MM2&13.23$\pm$2.13&(8.62$\pm$2.86)$\times$10$^{3}$&42$\pm$5&1.62$\pm$0.13&3.62&5.25$\times$10$^{6}$\\
~~~G358.93--0.03 MM2A&10.12$\pm$4.32&(3.16$\pm$1.36)$\times$10$^{3}$&27$\pm$3&1.52$\pm$0.18&3.52&2.58$\times$10$^{6}$\\
G358.93--0.03 MM3&19.21$\pm$4.57&(2.98$\pm$1.63)$\times$10$^{4}$&132$\pm$16&1.75$\pm$0.25&3.75& 5.28$\times$10$^{7}$\\	
G358.93--0.03 MM4&12.58$\pm$3.21&(3.56$\pm$1.02)$\times$10$^{3}$&38$\pm$13&1.63$\pm$0.15&3.63&3.26$\times$10$^{6}$\\
G358.93--0.03 MM5&14.47$\pm$2.63&(9.65$\pm$2.63)$\times$10$^{3}$&26$\pm$8&1.62$\pm$0.16&3.62&7.23$\times$10$^{6}$\\
G358.93--0.03 MM6&13.03$\pm$1.67&(7.23$\pm$1.02)$\times$10$^{3}$&22$\pm$6&1.58$\pm$0.23&3.58&4.27$\times$10$^{6}$\\
G358.93--0.03 MM7&11.53$\pm$2.48&(2.22$\pm$1.05)$\times$10$^{3}$&34$\pm$8&1.53$\pm$0.22&3.53&5.53$\times$10$^{6}$\\
G358.93--0.03 MM8&12.62$\pm$3.16&(4.63$\pm$1.12)$\times$10$^{3}$&24$\pm$4&1.58$\pm$0.27&3.58&1.26$\times$10$^{6}$\\
\hline 
\end{tabular}	
\end{adjustbox}
$\dagger$--The spectral index ($\alpha$) is calculated using the relation $\alpha = 2 + \beta$ \citep{be91}. \\
$\ddagger$--The density ($n$) is estimated using the equation $n = \frac{N(\mathrm{H}_2)}{2R}$, where $R$ is the radius of the core in cm \citep{liu21}.\\
\label{tab:dust}
\end{table*}

It is widely accepted that most stars, including high-mass stars ($M$ $\geq$ 8\(M_\odot\)), form in extremely dense clusters \citep{car00, la03, riv13}. This theory may also apply to the formation of our Sun \citep{ad10}. Investigating the physical and chemical properties of high-mass star-forming regions can provide valuable insight into the early stages of planetary system development. High-mass stars form within the compact ($D$ $\leq$ 0.1 pc) and dense ($n$ $\geq$ 10$^{5}$ cm$^{-3}$) cores of interstellar molecular clouds \citep{kur00, beu07, tan14, ya17}. As the central high-mass stars evolve slowly, their chemical compositions and physical properties (temperature, density, and luminosity) change \citep{beu07}. Previous observations reveal a sequence of stages, starting with high-mass starless cores (HMSCs), which are dense ($n$ $\geq$ 10$^{5}$ cm$^{-3}$), cold ($T$ $\sim$ 15--20 K), and undergo massive molecular condensation without evidence of star formation activity due to gravitational instability \citep{tan13}. The next phase is marked by the emergence of high-mass protostellar objects (HMPOs), characterized by protostar development within hot molecular cores (hereafter HMCs) at temperatures $T$ $\geq$ 100 K and gas densities $n$ $\geq$ 10$^{7}$ cm$^{-3}$ \citep{kur00, wil14, fon07}. As the protostar ignites, it emits ultraviolet (UV) photons, ionizes hydrogen, and heats the surrounding medium, thereby creating an H{\sc ii} region near the central star \citep{kur00}. The smaller H{\sc ii} regions are associated with younger massive stars, as ionization fronts expand supersonically owing to the stellar radiation pressure \citep{kur00}. In HMCs, the H{\sc ii} regions can be classified based on electron density and size: hypercompact H{\sc ii} regions (HC H{\sc ii}) have high densities ($n$ $\geq$ 10$^{6}$ cm$^{-3}$) and compact sizes ($R$ $\leq$ 0.05 pc), ultracompact H{\sc ii} regions (UC H{\sc ii}) have lower densities ($n$ $\geq$ 10$^{4}$ cm$^{-3}$) and sizes between 0.05 pc and 0.1 pc, and classical H{\sc ii} regions are larger than 0.1 pc \citep{wo89}.

\begin{figure*}
\centering
\includegraphics[width=0.9\textwidth]{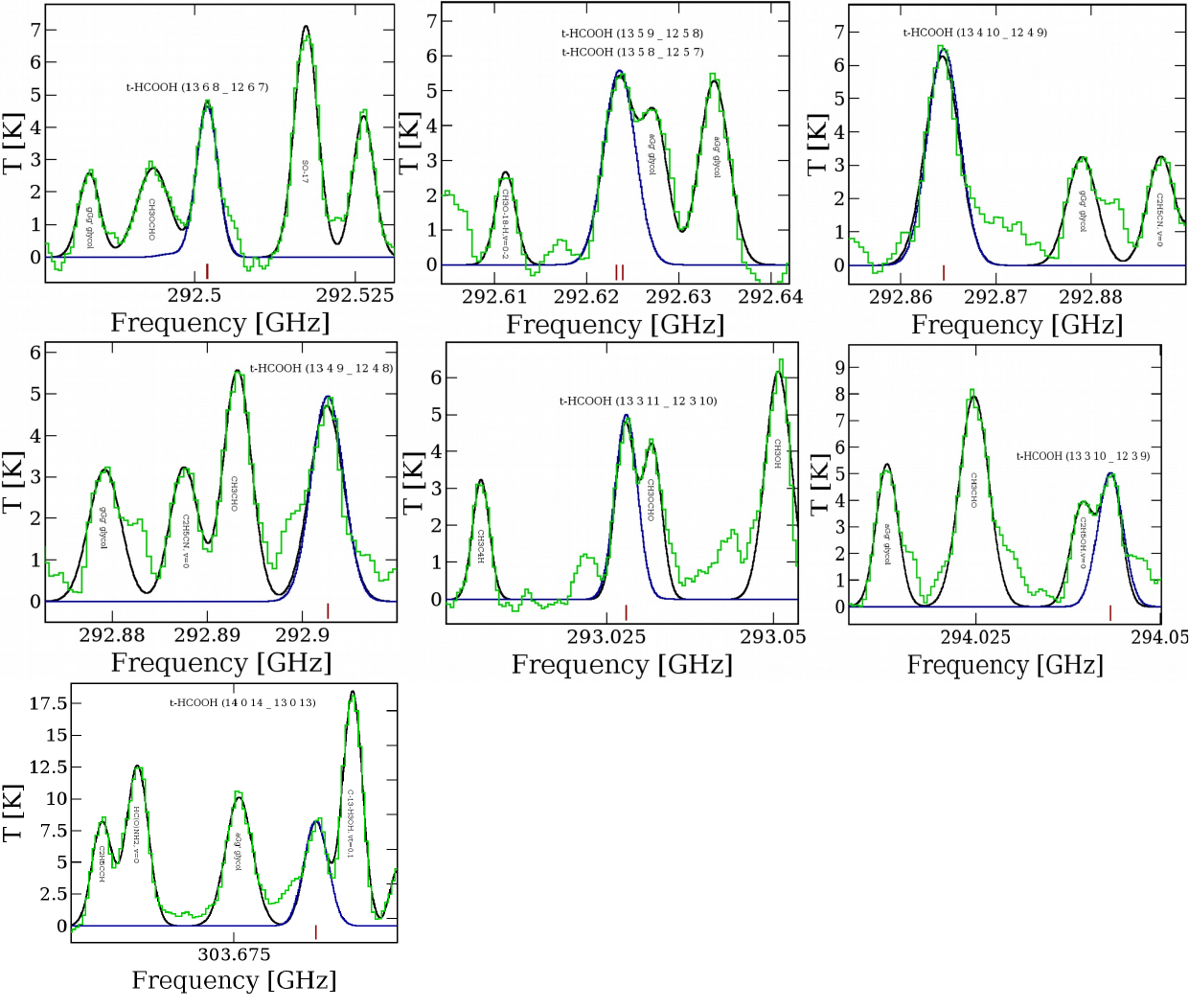}
\caption{Identified molecular emission lines $t$-HCOOH towards G358.93--0.03 MM1. The green lines are the observed spectra, and the blue lines are the LTE spectra of $t$-HCOOH. The black lines are the LTE spectra of the remaining molecules identified in the spectra of G358.93--0.03 MM1.}
\label{fig:spectra}
\end{figure*}

The massive star formation region G358.93--0.03 is located at a distance of 6.75$\,{}\,^{\,+0.37}_{\,-0.68}$\, kpc, whose gas mass and luminosity are 167$\pm$12\textup{M}$_{\odot}$ and $\sim$7.7$\times$10$^{3}$ \textup{L}$_{\odot}$, respectively \citep{re14, bro19}. This star formation region has eight dust continuum sources known as G358.93--0.03 MM1 to G358.93--0.03 MM8. Previously, \citet{mannaetal23} also found two more strong continuum sources, which are denoted as MM1A and MM2A (see Figure~1 in \citet{mannaetal23}). In this star formation region, \citet{bro19} classified G358.93--0.03 MM1 and G358.93--0.03 MM3 as hot molecular cores based on the detection of the emission lines of methyl cyanide (\ce{CH3CN}) and maser lines of methanol (\ce{CH3OH}). Recent studies of \cite{mannaetal23} and \cite{mannaetal24} show that G358.93--0.03 MM1 is a forest of complex molecules, but G358.93--0.03 MM3 is not molecularly rich. Previously, different chemical compounds were detected towards G358.93--0.03 MM1, such as cyanamide (\ce{NH2CN}) \citep{manna23b}, glycolaldehyde (\ce{CH2OHCHO}) \citep{mannaetal23}, ethylene glycol ((CH$_{2}$OH)$_{2}$) \citep{mannaetal24}, formamide (\ce{NH2CHO}) \citep{man24b}, isocyanic acid (HNCO) \citep{man24b}, \ce{NH2CH2COOH} precursor molecule methylamine (\ce{CH3NH2}) \citep{man24c}, ethylene oxide (c-\ce{C2H4O}) \citep{man26}, and acetaldehyde (\ce{CH3CHO}) \citep{man26}. After the detection of \ce{CH3NH2}, \cite{man24c} also searched the emission lines of \ce{NH2CH2COOH} without any detections. The upper-limit column densities of \ce{NH2CH2COOH} conformers I and II are $\leq$3.26$\times$10$^{15}$ cm$^{-2}$ and $\leq$1.20$\times$10$^{13}$ cm$^{-2}$, respectively \citep{man24c}. The identification of different types of molecular compounds indicates that this star formation region is an ideal source for studying various complex biomolecules, including acids.

This paper is presented in the following formats. In section~\ref{sec:obs}, we discuss the ALMA observation and data reduction procedures. The identification of HCOOH and its chemical formation routes are thoroughly described in sections~\ref{sec:re} and \ref{sec:dis}. The summary and conclusion are provided in section~\ref{sec:con}.

\section{Observation and data analysis}
\label{sec:obs}
The massive star-forming region G358.93--0.03 was observed using the Atacama Large Millimeter/submillimeter Array (ALMA) Band 7 receivers to study the high-mass protostellar accretion outburst (ID: 2019.1.00768.S, PI: Crystal Brogan). The observations were conducted on October 11th, 2019, with an on-source integration time of 756 seconds. The observed phase centre of G358.93--0.03 was ($\alpha,\delta$)$_{\rm J2000}$ = 17:43:10.000, -29:51:46.000. A total of 47 antennas were utilised, providing a minimum baseline of 14 metres and a maximum baseline of 2517 metres. The calibrator sources used during the observation were J1550+0527 (flux and bandpass calibrator) and J1744--3116 (phase calibrator). The observed frequency ranges for G358.93--0.03 were 290.51--292.39 GHz, 292.49--294.37 GHz, 302.62--304.49 GHz, and 304.14--306.01 GHz, with spectral and velocity resolutions of 1128.91 kHz and 0.96 km s$^{-1}$.

Data reduction and imaging were performed using the Common Astronomy Software Application (CASA 5.4.1) in conjunction with the ALMA data analysis pipeline \citep{mc07, team22}. At the time of data analysis, we applied the CASA pipeline tasks SETJY, hifa\_bandpassflag, and hifa\_flagdata for flux calibration, bandpass calibration, and flagging of disruptive antenna data. For flux calibration, the Perley-Butler 2017 flux calibrator model was utilised \citep{perl17}. After the initial data reduction, the target object G358.93--0.03 was split using the {MSTRANSFORM} task. Using the CASA task TCLEAN with the HOGBOM deconvolver, the dust continuum images of G358.93--0.03 were produced by utilising line-free channels in the observable frequency ranges. The UVCONTSUB task was used to subtract the background continuum emission from the UV plane of the calibrated data prior to constructing the spectral images. Using the TCLEAN task and the SPECMODE = CUBE parameter, spectral images of G358.93--0.03 were produced within the observable frequency ranges. To improve the RMS of the final images, we performed three rounds of phase-only self-calibration and one round of amplitude-only self-calibration using the GAINCAL and APPLYCAL tasks. We used the IMPBCOR task to correct the effect of the primary beam pattern in both the spectral and dust continuum images. The final dust continuum and line images were generated using CARTA \citep{cou21}.

\begin{table*}
\centering
\caption{Spectral line parameters of $t$-HCOOH. }	
\begin{adjustbox}{width=1.0\textwidth}
\begin{tabular}{ccccccccccccccccc}
\hline 
Observed frequency &Quantum number & $E_{up}$ & $A_{ij}$ &g$_{up}$&S$\mu^{2}$$^{\star}$ &FWHM&$\rm{\int T_{mb}dV}^\dagger$ &Remark\\
			
(GHz) &&(K)&(s$^{-1}$) & &(Debye$^{2}$) &(km s$^{-1}$)&(K km s$^{-1}$)& \\
\hline
~~292.501$^{*}$&13(6,8)--12(6,7)&212.63&2.23$\times$10$^{-4}$&27&20.67&3.53$\pm$0.21 &17.31$\pm$2.52 &Non blended \\
292.623&13(5,9)--12(5,8)&177.71&2.42$\times$10$^{-4}$&27&22.37&3.51$\pm$0.41 &16.71$\pm$3.75 &Non blended \\
292.623&13(5,8)--12(5,7)&177.71&2.42$\times$10$^{-4}$&27&22.37&3.51$\pm$0.32 &16.75$\pm$3.62&Non blended \\
292.864&13(4,10)--12(4,9)&149.16&2.57$\times$10$^{-4}$&27&23.77&3.54$\pm$0.27 &21.92$\pm$4.28 &Non blended \\
292.902&13(4,9)--12(4,8)&149.16&2.58$\times$10$^{-4}$&27&23.77&3.51$\pm$0.36 &15.51$\pm$2.74 &Non blended \\
293.027&13(3,11)--12(3,10)&126.96& 2.70$\times$10$^{-4}$&27&24.87&3.54$\pm$0.46 &15.78$\pm$2.63&Non blended \\
294.043&13(3,10)--12(3,9)&127.09&2.73$\times$10$^{-4}$&27&24.86&3.55$\pm$0.13 &15.14$\pm$2.43 &Non blended \\
303.687&14(0,14)--13(0,13)&110.93&3.17$\times$10$^{-4}$&29&28.17&3.52$\pm$0.19 &24.06$\pm$4.16 &Non blended\\			
\hline			
\end{tabular}	
\end{adjustbox}
\label{tab:MOLECULAR DATA}\\
{{*}}--Those transitions of $t$-HCOOH are contained double with frequency difference of $\leq$100 kHz. The second transition is not shown.\\
$\star$--The values of S$\mu^{2}$ is taken from the \href{https://splatalogue.online/#/home}{Splatalogue}.\\
$\dagger$--The values of $\rm{\int T_{mb}dV}$ is estimated by fitting the Gaussian model.\\
\end{table*}
 
\section{Results}
\label{sec:re}
\subsection{Dust continuum emission}
The dust continuum image of G358.93--0.03 at a frequency of 291.31 GHz is shown in Figure~\ref{fig:dustcontinuum}. The synthesized beam size and RMS noise level of the image are 0.41$^{\prime\prime}$$\times$0.36$^{\prime\prime}$ and 12.53 mJy, respectively. A total of eight main cores are identified, including two previously known cores associated with G358.93--0.03 MM1 and G358.93--0.03 MM2, hereafter referred to as G358.93--0.03 MM1A and G358.93--0.03 MM2A. To estimate the physical parameters of the detected cores, we performed two-dimensional Gaussian fitting using the task IMFIT. The derived parameters are summarized in Table~\ref{tab:cont}. Most of the dust cores are spatially resolved, except for G358.93--0.03 MM1A and G358.93--0.03 MM8. Previously, \citet{mannaetal23} also analyzed this dataset and estimated the \ce{H2} column densities of all cores in G358.93--0.03 based on the dust continuum emission.

\subsubsection{Estimation of physical parameters of dust cores}
We constructed the spectral energy distribution (SED) of the dust cores in G358.93--0.03 to derive key physical properties, including dust mass, luminosity, dust temperature ($T_{d}$), and the dust emissivity index ($\beta$). We also estimated the spectral index ($\alpha$) based on the derived values of $\beta$. Additionally, we derive the density ($n$) of all cores using the column density of H$_2$, with values taken from \citet{mannaetal23}. The SED incorporates flux density measurements from both ALMA bands 6 and 7, with the band 6 flux values adopted from \citet{manna23b}. The observed SEDs were fitted using the radiative transfer models developed by \citet{rob07}. During the fitting procedure, the visual extinction ($A_{v}$) was constrained to a range of 0--30 mag, consistent with the values reported by \citet{fu16}, and the source distance was fixed at 6.75 kpc. Details of the radiative transfer fitting methodology can be found in \citet{san14} and \citet{min21}. The resulting SED fits and the derived dust properties of all cores are presented in Figure~\ref{fig:SED} and Table~\ref{tab:dust}. We find that the dust temperatures of G358.93--0.03 MM1 and MM3 are higher than those of the other cores, consistent with their identification as HMCs. Previously, \citet{bro19} derived a luminosity of 7.7$\times$10$^{3}$ \textup{L}$_{\odot}$ for G358.93--0.03, but our analysis shows that the hot cores MM1 and MM3 exhibit relatively higher luminosities. We find that the $\beta$ varies across the cores, ranging from 1.53 to 1.72. These values are consistent with typical ISM dust, where $\beta \sim 1.6$, corresponding to maximum grain sizes between approximately 100 $\AA$ and 0.3 $\mu$m \citep{wei01}. We also find that the densities of G358.93--0.03 MM1 and G358.93--0.03 MM3 are 1.08$\times$10$^{8}$ and 5.28$\times$10$^{7}$ cm$^{-3}$, respectively. The other dust cores show relatively lower densities, on the order of $\sim$10$^{6}$ cm$^{-3}$. The high densities observed in G358.93--0.03 MM1 and G358.93--0.03 MM3 further support their classification as hot core candidates.

\begin{figure}
\centering
\includegraphics[width=0.5\textwidth]{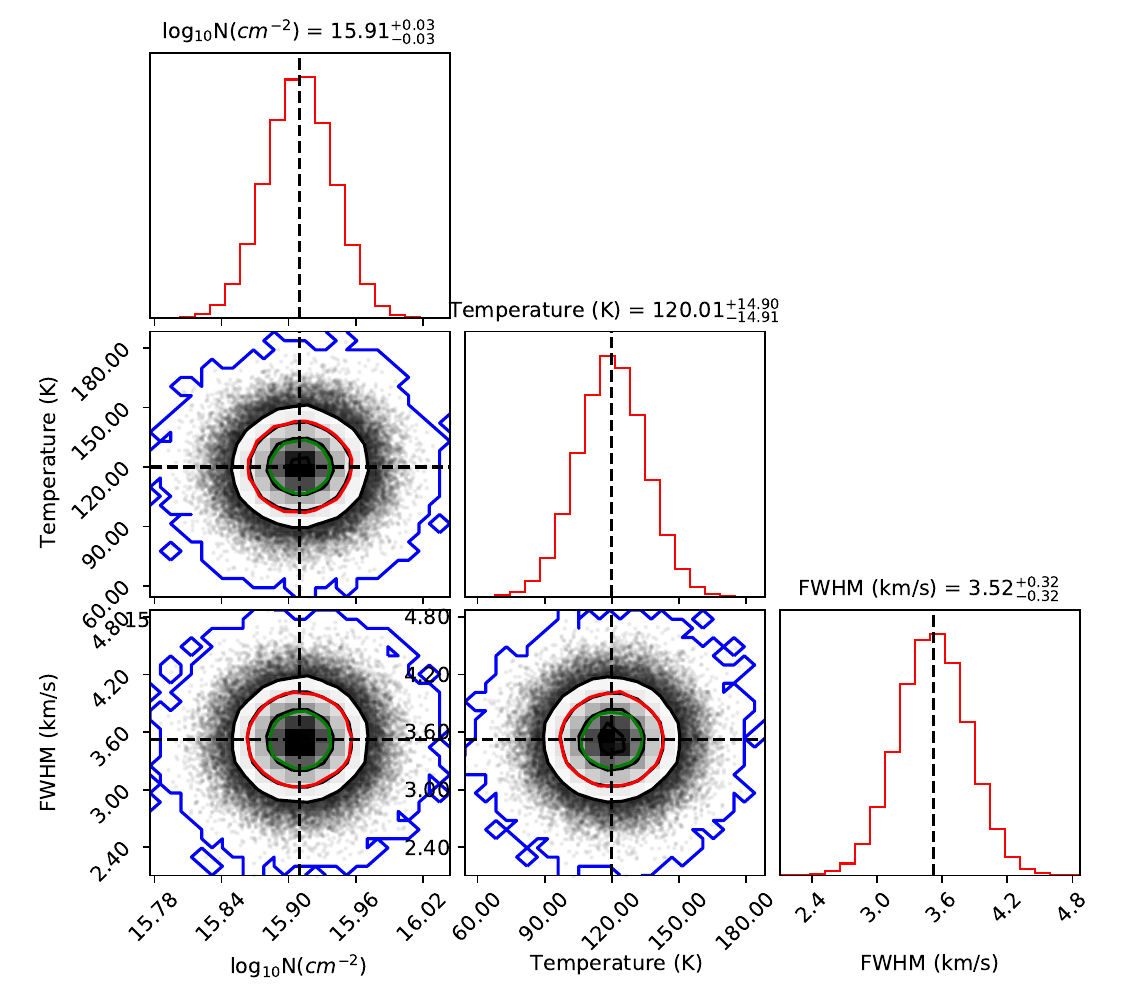}
\caption{Corner plots showing the covariances of the posterior probability distributions of the column density (log$_{10}$($N$)) in cm$^{-2}$, excitation temperature in K, and FWHM in km s$^{-1}$ of $t$-HCOOH.}
\label{fig:columncor}
\end{figure}

\subsection{Line emissions from G358.93--0.03}
We found that the hot cores G358.93--0.03 MM1 and G358.93--0.03 MM3 emit molecular line emissions based on the spectral images of G358.93--0.03. We also noticed that the other continuum cores emit no molecular lines. The synthesized beam sizes for the spectral images are 0.42$^{\prime\prime}$$\times$0.37$^{\prime\prime}$, 0.43$^{\prime\prime}$$\times$0.37$^{\prime\prime}$, 0.41$^{\prime\prime}$$\times$0.37$^{\prime\prime}$, and 0.41 $^{\prime\prime}$$\times$0.36$^{\prime\prime}$ at frequency ranges of 290.51--292.39 GHz, 292.49--294.37 GHz, 302.62--304.49 GHz, and 304.14--306.01 GHz, respectively. We extracted line spectra from G358.93--0.03 MM1 and G358.93--0.03 MM3 using a 0.52$^{\prime\prime}$ diameter circular region, which encompasses the line-emitting regions of both hot cores. The extracted spectra are shown in Figure~2 of \cite{mannaetal23}. Based on the molecular spectra, we see that G358.93--0.03 MM1 is more chemically rich than G358.93--0.03 MM3. We observed some emission lines of \ce{CH3OH}, \ce{H2CO}, and CS showing the outflow nature. The systematic velocities ($V_{LSR}$) of the molecular spectra of G358.93--0.03 MM1 and G358.93--0.03 MM3 are --16.5 km s$^{-1}$ and --18.5 km s$^{-1}$, respectively \citep{bro19}.

\begin{figure}
\centering
\includegraphics[width=0.48\textwidth]{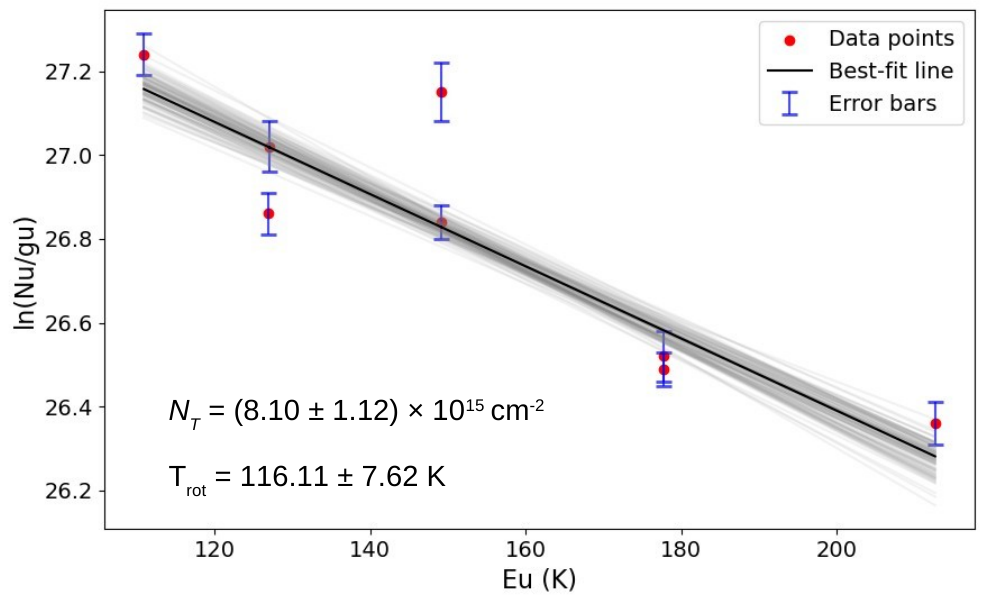}
\caption{Rotational diagram of $t$-HCOOH for the estimation of the total column density ($N_{T}$) and rotational temperature ($T_{rot}$) of $t$-HCOOH using the MCMC approach. The grey lines are the random draws from the fit posteriors, and the solid black line in the rotational diagram represents the best-fit straight line.}
\label{fig:rotd}
\end{figure}

Following spectrum extraction, we used the CASSIS software to identify molecules and analyse spectrums using the local thermodynamic equilibrium (LTE) model with the help of molecular spectroscopic databases from the Jet Propulsion Laboratory (JPL) and Cologne Database for Molecular Spectroscopy (CDMS) \citep{vas15, pic98, mu05}. We determine the excitation temperatures and column densities of the identified molecules using the LTE-modelled spectra. CASSIS uses the following formula to determine the brightness temperature ($T_{b}$) of the molecular spectra:

\begin{equation}
T_{b} = T_{C}e^{-\tau} +(1-e^{-\tau})(J_{\nu}(T_{ex})- J_{\nu}(CMB))
\label{eq:tb}
\end{equation}
In equation~\ref{eq:tb}, $J_{\nu}(T)$ $ = (h\nu/k)\times1/(e^{h\nu/k T}-1)$ is the radiation temperature, $\tau$ is the optical depth, $T_{C}$ is the continuum temperature, and CMB is the cosmic microwave background at 2.7 K. We fitted the LTE-modelled spectra of molecules over the observed spectra using the Markov chain Monte Carlo (MCMC) technique in the CASSIS. The first step of the MCMC method is to choose an initial point ($X_0$) at random from the four-dimensional parameter space. After that, it iteratively creates neighbouring points ($X_1$) using a dynamically modified step size. After evaluating each new point's $\chi^{2}$ value, the new point is accepted if the ratio $p$ = $\chi^{2}$($X_0$)/$\chi^{2}$($X_1$) is greater than 1. Even so, the method can avoid local minima and fully search the parameter space if $p$ is smaller than 1 because the new point may still be accepted with a given probability. In our MCMC study, we used 2000 walkers, evenly distributed throughout defined parameter ranges. For absolute convergence, we ran the chains for 22000 steps. In order to achieve the best fit, this method allows us to simultaneously change all of the parameters, such as column density, excitation temperature, full width at half maximum (FWHM), and V$_{LSR}$. CASSIS calculated the partition functions of the identified molecules using the following expression:
	
\begin{equation}
Q(T) = \sum_{i} g_{i} \times exp(-E_{i}/kT)
\label{eq:stat}
\end{equation}
In equation~\ref{eq:stat}, $g_{i}$ and $E_{i}$ represent the statistical weight and energy of the $i$th energy level, respectively. We employed a source size of $\sim$0.5$^{\prime\prime}$ (the half-maximum diameter of our beam size) for LTE spectral modelling. We evaluated the column density, excitation temperature, and FWHM of the detected molecule following spectrum analysis using the LTE models. Additionally, we use the column densities of identified molecules and divide them by the \ce{H2} column density to estimate the fractional abundances of detected molecules. After spectral analysis, we found the rotational emission lines of $t$-HCOOH, which are discussed in the next section.

\begin{figure}
\centering
\includegraphics[width=0.5\textwidth]{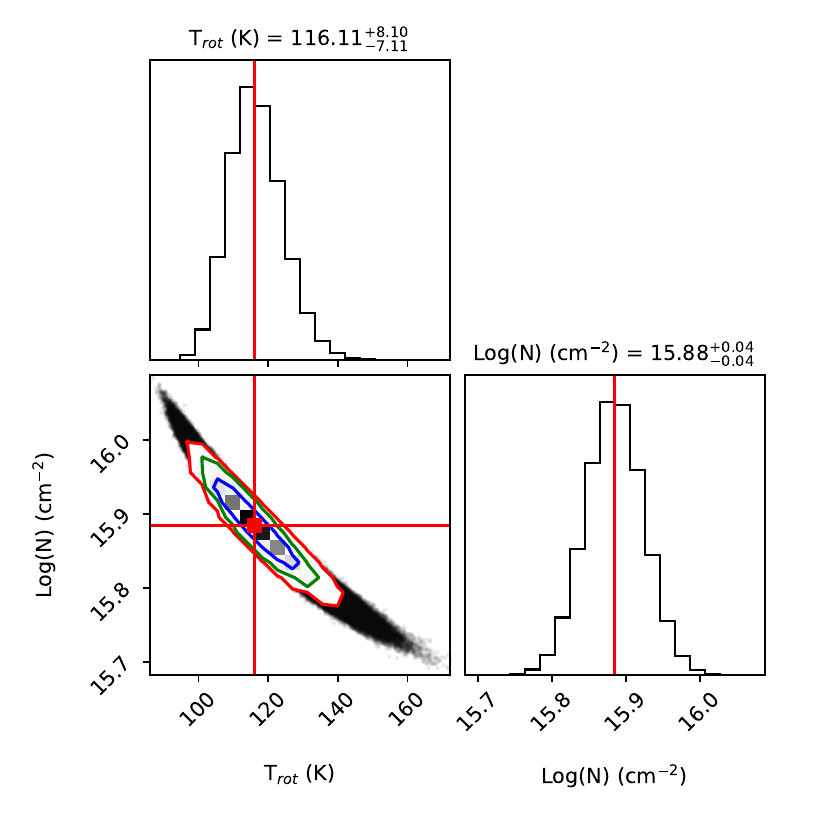}
\caption{Marginalized posterior probability distribution (corner plot) of column density and rotational temperature of $t$-HCOOH using the MCMC approach.}
\label{fig:cornerrotd}
\end{figure}

\begin{table}
\centering
\caption{Derived emitting regions of $t$-HCOOH.}
\begin{adjustbox}{width=0.48\textwidth}
\begin{tabular}{ccccccccccccccccc}
\hline 
Observed frequency&Quantum number&$E_{u}$ &$\theta_{S}$\\
(GHz)            &         & (K)     &($^{\prime\prime}$) \\
\hline
292.501&13(6,8)--12(6,7)&212.63&0.41 \\
292.623&13(5,9)--12(5,8)&177.71&0.42 \\
292.623&13(5,8)--12(5,7)&177.71&0.42 \\
292.864&13(4,10)--12(4,9)&149.16&0.40\\
292.902&13(4,9)--12(4,8)&149.16&0.42 \\
293.027&13(3,11)--12(3,10)&126.96&0.43 \\
294.043&13(3,10)--12(3,9)&127.09&0.42 \\
303.687&14(0,14)--13(0,13)&110.93&0.41\\
\hline
\end{tabular}	
\end{adjustbox}
\label{tab:emitting region}
\end{table}

\subsubsection{$t$-HCOOH towards G358.93--0.03 MM1}
Following spectral analysis, we identified a total of 8 transition lines of $t$-HCOOH towards G358.93--0.03 MM1, where upper-state energies ($E_{up}$) change between 110.93 K and 212.63 K. This is the first time this molecule has been found in G358.93--0.03 MM1, as far as we know. After spectral analysis, we found that all detected emission lines of $t$-HCOOH within the studied frequency ranges are non-blended and detected above 3.5$\sigma$ significance. We also observed that $J$ = 13(5,9)--12(5,8) and $J$ = 13(5,8)--12(5,7) transitions of $t$-HCOOH are found in the single spectral line because those transitions are located in the short frequency span. Figure~\ref{fig:spectra} and Table~\ref{tab:MOLECULAR DATA} display the LTE-fitted spectral lines and line parameters of $t$-HCOOH. For $t$-HCOOH, the column density ($N$), excitation temperature ($T_{ex}$), and FWHM are $(8.13\pm0.72)\times10^{15}$ cm$^{-2}$, $120\pm15$ K, and $3.52\pm0.32$ km s$^{-1}$, respectively, based on the posterior probability distribution shown in Figure~\ref{fig:columncor}. The fractional abundance of $t$-HCOOH with respect to \ce{H2} is $(2.62\pm0.29)\times10^{-9}$, where the column density of \ce{H2} towards G358.93--0.03 MM1 is $(3.1\pm0.2)\times10^{24}$ cm$^{-2}$ \citep{mannaetal23}. During spectral analysis, we also detected the emission lines of \ce{CH3OH} and \ce{H2CO} in the spectra of G358.93--0.03 MM1, and we discuss the detailed physical and chemical properties of both molecules in a separate paper. We noticed that emission lines of \ce{H2CO} show an outflow nature. In the case of \ce{CH3OH}, we see the lower $E_{up}$ ($\leq$50 K) of \ce{CH3OH} lines show outflows nature, but higher $E_{up}$ ($\geq$70 K) \ce{CH3OH} lines show compact nature, meaning those lines are emitted from the warm inner region of G358.93--0.03 MM1. To confirm the outflow nature of both \ce{CH3OH} and \ce{H2CO}, we constructed channel maps for each molecule using the transitions $J$ = 2(1,-0)--2(0,+0) for \ce{CH3OH} and $J$ = 4(0,4)--3(0,3) for \ce{H2CO}. The channel maps, along with the corresponding spectra of \ce{CH3OH} and \ce{H2CO}, are presented in Figures~\ref{fig:CH3OHCHANNEL} and \ref{fig:H2COCHANNEL}, respectively. The column density and excitation temperature of \ce{H2CO} are $(7.02\pm0.35)\times10^{15}$ cm$^{-2}$ and $90\pm6$ K. The column density and excitation temperature of \ce{CH3OH}, which exhibit the outflow nature, are $(8.52\pm0.46)\times10^{16}$ cm$^{-2}$ and $80\pm7$ K. The column density and excitation temperature of \ce{CH3OH}, which emitted from the warm inner region, are $(5.21\pm0.22)\times10^{17}$ cm$^{-2}$ and $152\pm16$ K. The column density ratios of $t$-HCOOH/\ce{CH3OH} and $t$-HCOOH/\ce{H2CO} are $(1.56\pm0.12)\times10^{-2}$ and $(1.16\pm0.12)$. To estimate the $t$-HCOOH/\ce{CH3OH} value, we use the column density of \ce{CH3OH}, which is emitted from the warm inner regions. After the detection of $t$-HCOOH, we also searched for the emission lines of $c$-HCOOH, but we observed that all detected lines of $c$-HCOOH are blended. So, we estimated the upper-limit column density of $c$-HCOOH to be $\leq(7.25\pm0.83)\times10^{14}$ cm$^{-2}$.

\begin{table}
\centering
\caption{Initial fractional abundances of \ce{H2} and atomic elements at the start of the collapse phase.}
\begin{adjustbox}{width=0.27\textwidth}
\begin{tabular}{ccc}
\hline 
Species&  & Abundance\\
\hline
\ce{H2}&&4.99$\times$10$^{-1}$ \\
H & &2.00$\times$10$^{-3}$ \\
He & &9.00$\times$10$^{-2}$ \\
N & &7.50$\times$10$^{-5}$ \\
C & & 1.40$\times$10$^{-4}$ \\
O & & 3.20$\times$10$^{-4}$ \\
S & &8.00$\times$10$^{-8}$\\
Mg & &7.00$\times$10$^{-9}$\\
Na & &2.00$\times$10$^{-8}$\\
P & &3.00$\times$10$^{-9}$\\
Si & &8.00$\times$10$^{-9}$\\
Fe & &3.00$\times$10$^{-9}$\\
Cl & &4.00$\times$10$^{-9}$\\
\hline
\end{tabular}	
\end{adjustbox}
\label{tab:mod}
\end{table}

\begin{figure*}
\centering
\includegraphics[width=0.9\textwidth]{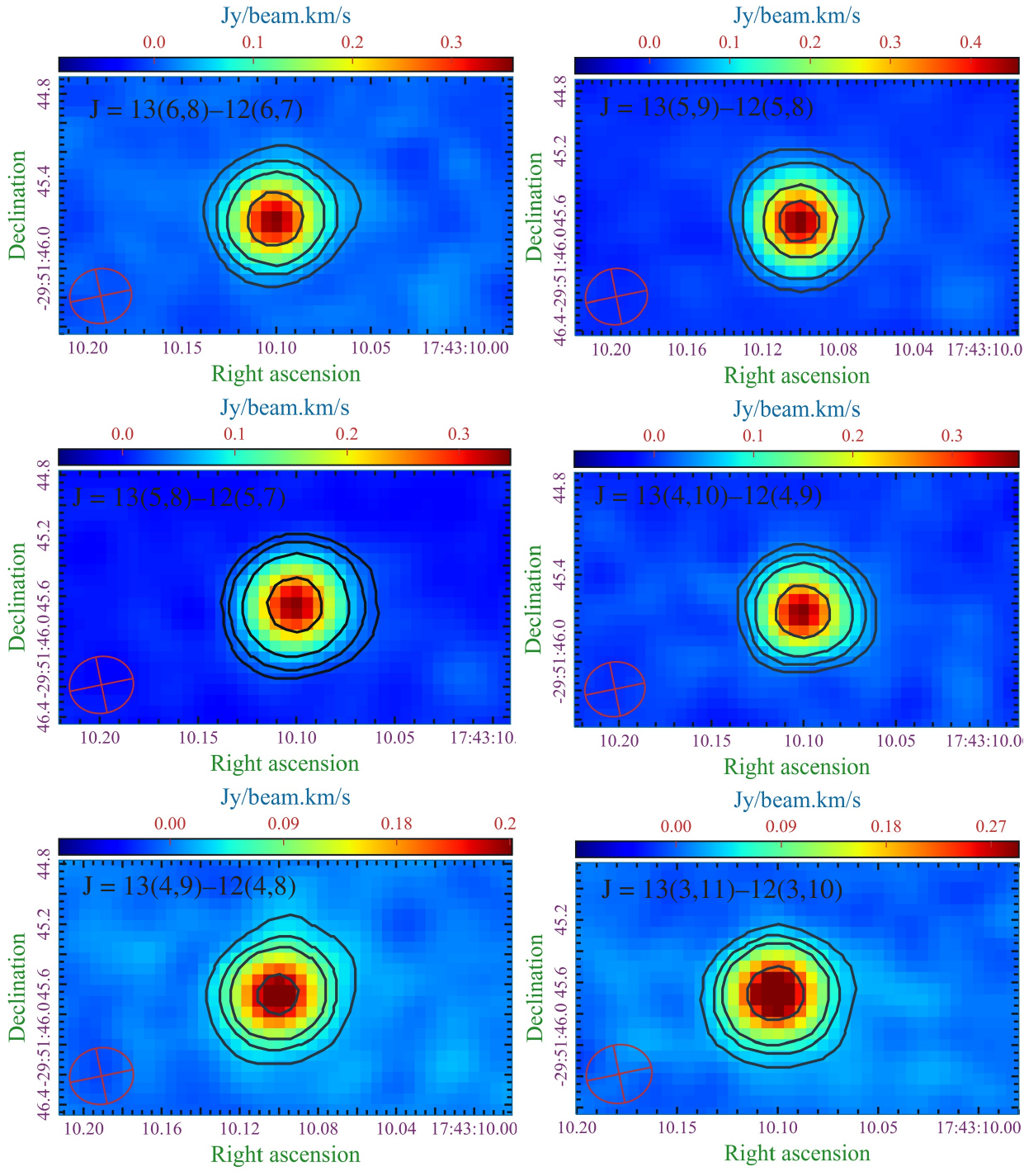}
\includegraphics[width=0.9\textwidth]{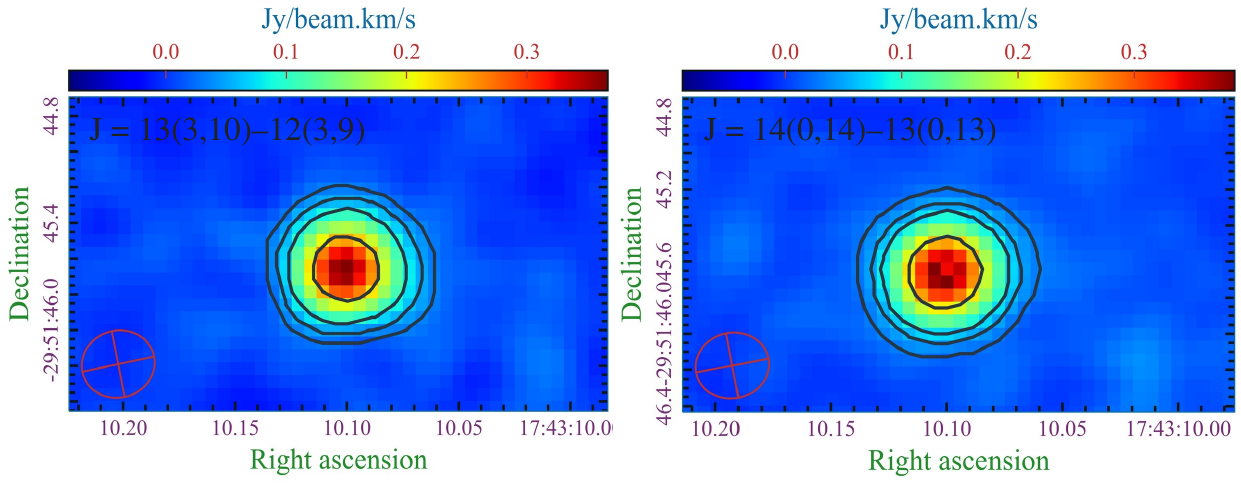}
\caption{Integrated intensity images (moment zero maps) of the detected transitions of $t$-HCOOH towards G358.93--0.03 MM1. The black contours are the 291.31 GHz continuum emission, and the contour levels are at 20\%, 40\%, 60\%, and 80\% of the peak flux. The red circles represent the synthesized beam of the intensity images.}
\label{fig:map}
\end{figure*}

\begin{figure*}
\centering
\includegraphics[width=0.88\textwidth]{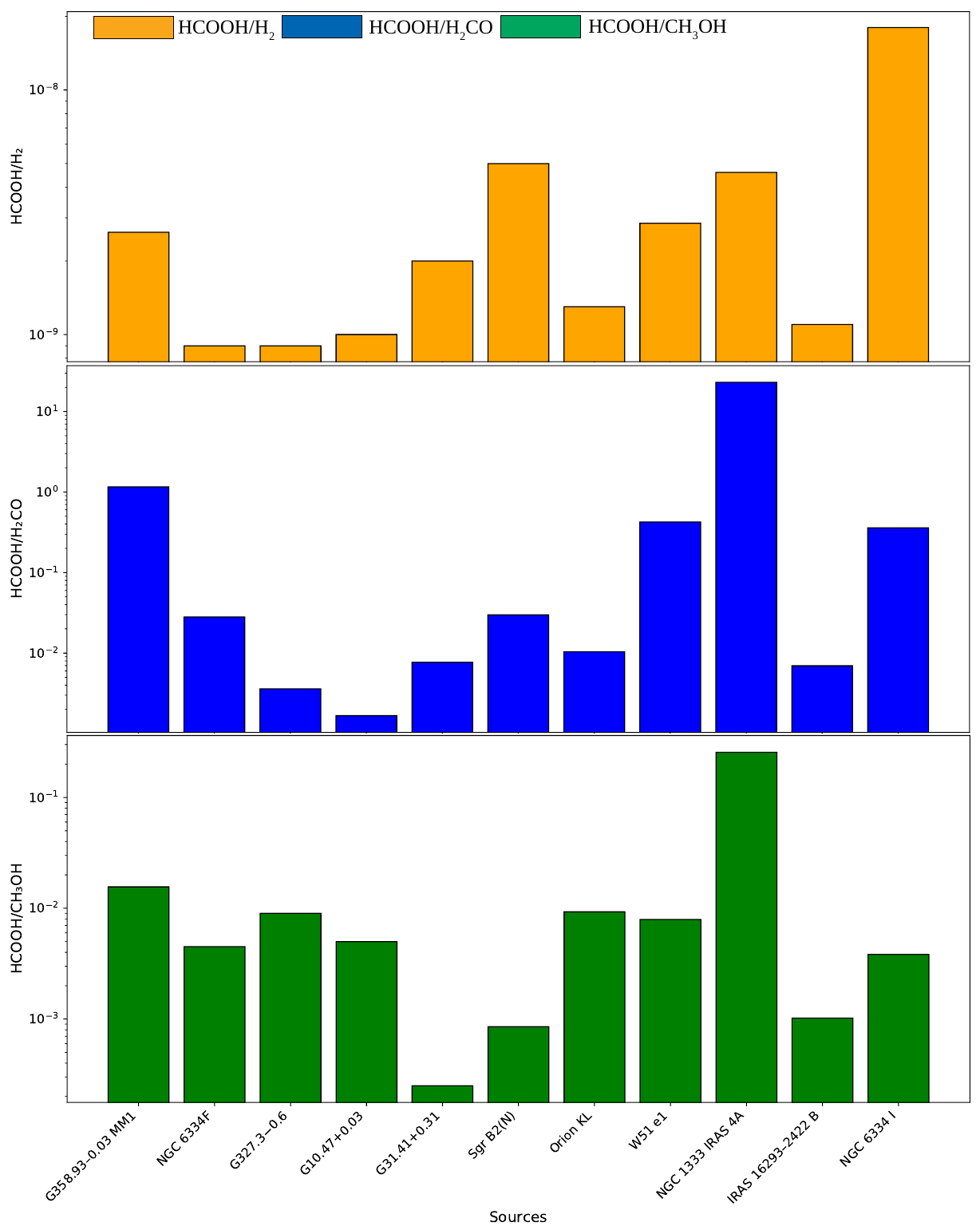}
\caption{Comparison of the abundances of HCOOH relative to \ce{H2} (upper panel), \ce{H2CO} (middle panel), and \ce{CH3OH} (lower panel) towards G358.93--0.03 MM1, with different hot cores (i.e., NGC 6334F, G327.3--0.6, G10.47+0.03, G31.41+0.31, Sgr B2(N), Orion KL, W51 e1, NGC 6334 I), and hot corinos (i.e., NGC 1333 IRAS 4A, IRAS 16293--2422 B).}
\label{fig:bar}
\end{figure*}

\begin{figure*}
	\centering
	\includegraphics[width=1.0\textwidth]{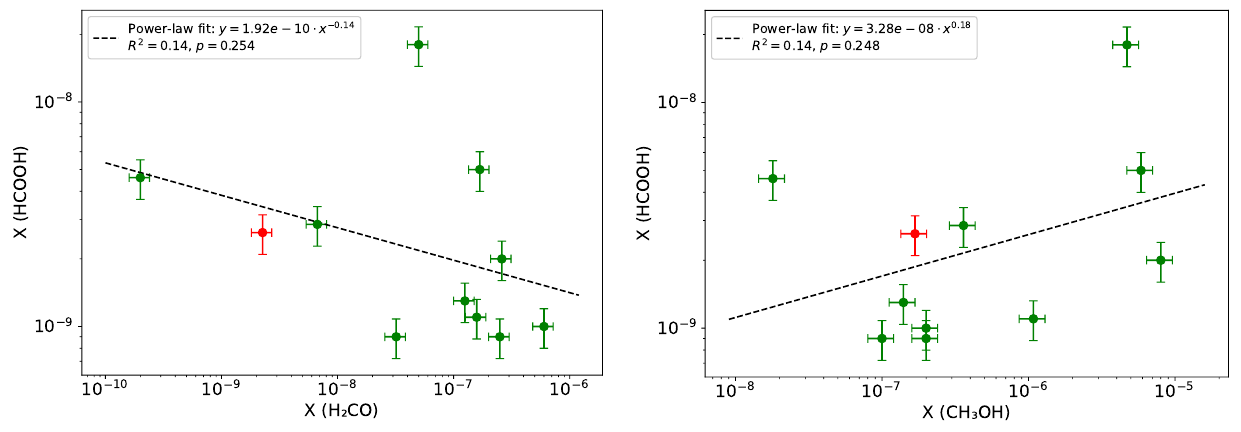}
	\caption{Relationship between the molecular abundances of HCOOH, \ce{CH3OH}, and \ce{H2CO} across different hot cores. The red data points correspond to measurements in G358.93--0.03 MM1, while the black dashed lines indicate the best-fit power-law.}
	\label{fig:comp}
\end{figure*}

\subsubsection{Estimation of rotational temperature: Rotational diagram}
We have computed a rotational diagram based on the eight non-contaminated transitions of $t$-HCOOH in order to determine the rotational temperature ($T_{rot}$) and total column density ($N_{T}$). The rotational diagram is essential for confirming the estimated temperature and column density derived from the LTE modelling. Our initial assumption was that the detected $t$-HCOOH emission lines were optically thin and populated in an environment of LTE. The column density equation for optically thin molecular lines is expressed as \citep{gold99}:

\begin{equation}
{N_u^{thin}}=\frac{3{g_u}k_B}{8\pi^{3}\nu S\mu^{2}}\int{T_{mb}dV}
\label{eq:rotd1}
\end{equation}
In equation~\ref{eq:rotd1}, $g_u$ is the degeneracy, $k_B$ is the Boltzmann constant, $\mu$ is the electric dipole moment of the molecule, $\nu$ is the rest frequency of the molecule, $S$ is the strength of the molecule, and $\int T_{mb}dV$ is the integrated intensity of the molecular line in km s$^{-1}$. Under the LTE conditions, the total column density of the molecule is expressed as

\begin{equation}
\frac{N_u^{thin}}{g_u} = \frac{N_{total}}{Q(T_{rot})}\exp\left(\frac{-E_u}{k_BT_{rot}}\right)
\label{eq:rotd2}
\end{equation}
In equation~\ref{eq:rotd2}, $T_{rot}$ represents the rotational temperature of the molecule, and ${Q(T_{rot})}$ is the partition function at the estimated rotational temperature. The value of ${Q(T_{rot})}$ of $t$-HCOOH are 8901.55 ($T_{rot}$ = 300 K), 5779.97 ($T_{rot}$ = 225 K), 3145.85 ($T_{rot}$ = 150 K), and 1112.92 ($T_{rot}$ = 75 K) \citep{mu05}. After rearranging, the equation~\ref{eq:rotd2} can be written as,

\begin{equation}
ln\left(\frac{N_u^{thin}}{g_u}\right) = ln(N)-ln(Q)-\left(\frac{E_u}{k_BT_{rot}}\right)
\label{eq:rotd3}
\end{equation}
Equation~\ref{eq:rotd3} showed that the $E_{u}$ of the molecule and $\ln(N_{u}/g_{u})$ had a linear connection. Using equation~\ref{eq:rotd2}, the value $\ln(N_{u}/g_{u})$ was estimated. Equation~\ref{eq:rotd3} suggests that the spectral parameters for different transition lines of $t$-HCOOH should be fitted with a straight line, where the slope is inversely proportional to the rotational temperature $T_{rot}$. After fitting a Gaussian model across the non-blended spectra of $t$-HCOOH, we determined the spectral line parameters of $t$-HCOOH for the rotational diagram. For fitting the equation~\ref{eq:rotd3} over the $\ln(N_{u}/g_{u})$ plotted against $E_{u}$, we used the MCMC approach using the Python module {\tt emcee} \citep{for13}. After fitting equation~\ref{eq:rotd3} using the MCMC method, the {\tt emcee} module generates posterior probability distributions of $N_{T}$ and $T_{rot}$. The rotational diagram and marginalized posterior probability distribution are shown in Figures~\ref{fig:rotd} and \ref{fig:cornerrotd}. Based on the rotational diagram, we found the value of total column density ($N_{T}$) and $T_{rot}$ are $(8.10\pm1.12)\times10^{15}$ cm$^{-2}$ and 116.11 $\pm$ 7.62 K. We noticed that the values of $T_{rot}$ and $T_{ex}$ are nearly the same ($T_{rot} \sim T_{ex}$), which indicates that $t$-HCOOH is in LTE conditions in G358.93--0.03 MM1.

\subsubsection{Analysis of HCOOH towards G358.93--0.03 MM3}
Using the LTE-modelled spectra, we also searched the emission lines of $t$-HCOOH and $c$-HCOOH towards G358.93--0.03 MM3, but we did not detect those molecules. The upper-limit column densities of $t$-HCOOH and $c$-HCOOH are $\leq(2.53\pm0.62)\times10^{13}$ cm$^{-2}$ and $\leq(3.28\pm0.55)\times10^{12}$ cm$^{-2}$, respectively. The upper limits of their fractional abundance are $\leq(7.23\pm2.32)\times10^{-11}$ and  $\leq(9.37\pm2.45)\times10^{-12}$, respectively, where the value of $N(H_{2})$ towards G358.93--0.03 MM3 is $(3.5 \pm 0.7)\times10^{23}$ cm$^{-2}$ \citep{mannaetal24}.

\subsection{Spatial distributions of $t$-HCOOH}
Using the CASA task IMMOMENTS, we generated the integrated emission maps of the detected $t$-HCOOH emission lines toward G358.93--0.03 MM1 (Figure~\ref{fig:map}). These maps are overlaid with the 291.31 GHz dust continuum to compare peak positions, which are found to be spatially coincident. The compact morphology suggests that $t$-HCOOH emission originates from the dense and warm inner region of hot core G358.93--0.03 MM1. We estimated the emitting region sizes ($\theta_{S}$) by fitting 2D Gaussians with the IMFIT task, and the results are summarized in Table~\ref{tab:emitting region}. The derived $\theta_{S}$ values (0.40$^{\prime\prime}$--0.43$^{\prime\prime}$) are below than the synthesized beam sizes ($0.43^{\prime\prime}\times0.36^{\prime\prime}$ to $0.45^{\prime\prime}\times0.38^{\prime\prime}$), indicating that the $t$-HCOOH emission is unresolved. Therefore, we do not draw any conclusions about the chemical morphology of $t$-HCOOH based on its spatial distribution. Thus, its spatial morphology remains uncertain and would benefit from higher-resolution observations.

\begin{figure*}
	\centering
	\includegraphics[width=0.5\textwidth]{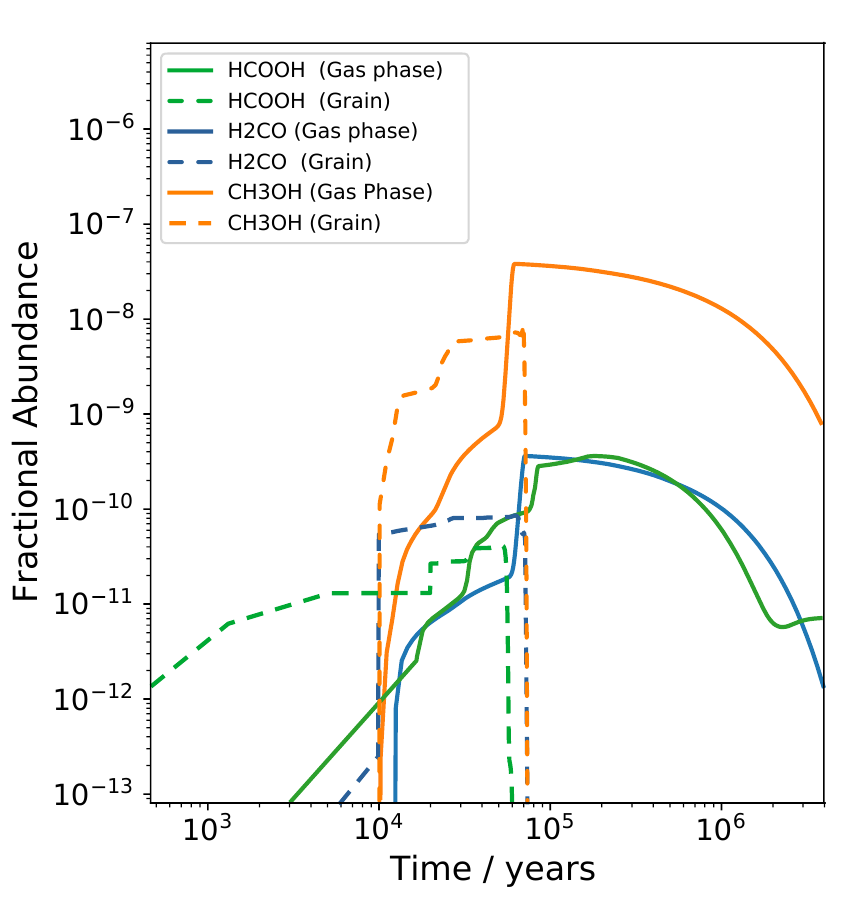}\includegraphics[width=0.5\textwidth]{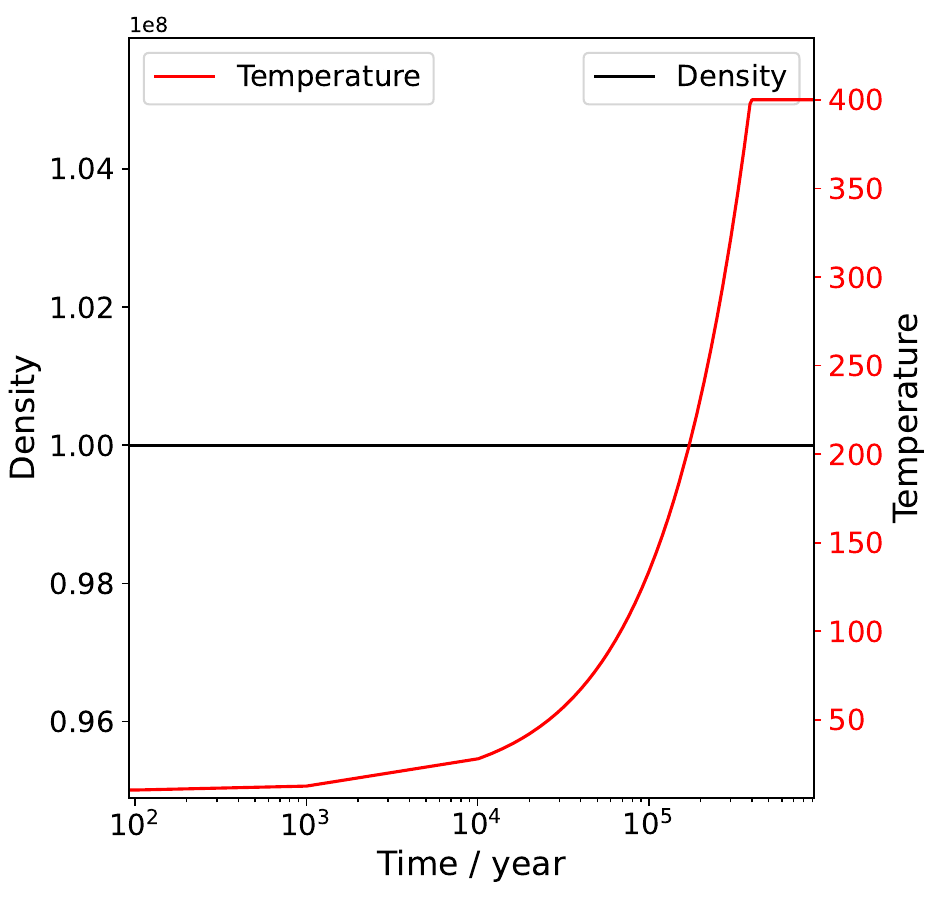}
	\caption{Time-dependent fractional abundances of HCOOH, \ce{H2CO}, and \ce{CH3OH} (left figure). Gas-phase species are represented by solid lines, while their grain (including both surface and mantle) counterparts are depicted with dashed lines of the same colour. The right-hand figure shows the temperature profile during the warm-up stage.}
	\label{fig:model}
\end{figure*}

\section{Discussions}
\label{sec:dis}
\subsection{Comparison and correlation analysis of HCOOH abundances with those in other sources}
Here, we compared the derived abundances of HCOOH relative to \ce{H2}, \ce{H2CO}, and \ce{CH3OH} between G358.93--0.03 MM1 and other sources. A comparison is made between the abundances of HCOOH found in G358.93--0.03 MM1 and various hot core and hot corino sources, such as NGC 6334 F \citep{nu98, zer12}, G327.3--0.6 \citep{nu98, wr06}, G10.47+0.03 \citep{nu98, rof11}, G31.41+0.31 \citep{gar22, min23, lo24}, Sgr B2(N) \citep{bel13}, Orion KL \citep{peng19, fav, zer12}, W51 e1 \citep{li01, ro16}, NGC 1333 IRAS 4A \citep{bot04, mar04, qu24}, IRAS 16293--2422 B \citep{mani20, per18}, and NGC 6334 I \citep{zer12}. Figure~\ref{fig:bar} shows a bar diagram comparing the abundances of HCOOH towards G358.93--0.03 MM1 and other sources. It is evident that different hot cores and hot corinos have varying HCOOH abundances, which suggests that their chemical processes vary.

After comparison, we found the abundance of HCOOH relative to \ce{H2} towards G358.93--0.03 MM1 is higher than most of the sources except for NGC 6334 I and NGC 1333 IRAS 4A. The abundance of HCOOH relative to \ce{H2} towards G358.93--0.03 MM1 is comparable to that of G31.41+0.31 and slightly lower than that of W51 e1. Likewise, the HCOOH/\ce{CH3OH} ratio for G358.93--0.03 MM1 is substantially higher than that of the majority of other sources, such as NGC 6334 F, G10.47+0.03, and W51 e1. The HCOOH/\ce{CH3OH} ratio for G358.93--0.03 MM1 is significantly greater than that of G31.41+0.31 and Sgr B2 (N), suggesting that G358.93--0.03 MM1 has a particularly high HCOOH abundance relative to \ce{CH3OH}. We observed that the NGC 1333 IRAS 4A exhibited the highest HCOOH/\ce{CH3OH} ratio compared to all other sources, indicating that the region may produce more HCOOH because of various physical or chemical circumstances. The HCOOH/\ce{H2CO} ratio towards G358.93--0.03 MM1 is remarkably high compared to most sources. The HCOOH/\ce{H2CO} ratio towards G358.93--0.03 MM1 is significantly greater than those found in NGC 6334 F, G10.47+0.03, and G327.3--0.6. The only source with a much higher ratio is NGC 1333 IRAS 4A, whereas intermediate values are observed in W51 e1 and NGC 6334 I. The high \ce{HCOOH/H2CO} ratio in G358.93--0.03 MM1 suggests that the formation pathway of HCOOH may be more efficient than \ce{H2CO}, possibly due to variations in physical conditions such as temperature, density, or UV radiation exposure. Overall, these results indicate that G358.93--0.03 MM1 exhibits a relatively high abundance of HCOOH relative to \ce{H2}, \ce{CH3OH}, and \ce{H2CO}, suggesting that this source is chemically rich in HCOOH relative to other interstellar environments. The enhanced HCOOH abundance may be linked to specific astrochemical pathways such as ice mantle processing, grain-surface reactions, or gas-phase chemistry influenced by shock activity or radiation fields. Further investigations of the physical conditions of G358.93--0.03 MM1 could provide insights into the dominant formation mechanisms of HCOOH in this region.
		
To explore the chemical relationship between HCOOH and its potential precursors, we examined the correlation between the observed abundances of HCOOH and two commonly proposed parent species, \ce{H2CO} and \ce{CH3OH}, across a sample of well-known hot cores. The resulting molecular correlations are presented in Figure~\ref{fig:comp}.
	
In the first analysis (see left panel in Figure~\ref{fig:comp}), we plotted the fractional abundance of HCOOH against that of \ce{H2CO} and performed a linear regression to assess the strength and nature of their correlation. The power-law fitting yields a coefficient of determination ($R^{2}$) of 0.14 and a $p$-value of 0.254. This result indicates a very weak and negative correlation between \ce{H2CO} and HCOOH abundances, suggesting that the two molecules do not scale proportionally across different sources. From a chemical standpoint, this weak and negative correlation implies that while \ce{H2CO} might participate in reaction pathways that produce HCOOH (e.g., through hydrogenation or radical chemistry on grain surfaces), its overall abundance is not a strong predictor of the amount of HCOOH present in these environments.

In the second analysis (see right panel in Figure~\ref{fig:comp}), we examined the relationship between \ce{CH3OH} and HCOOH abundances. Methanol is considered a more chemically relevant precursor to HCOOH, particularly through grain-surface chemistry. The regression between \ce{CH3OH} and HCOOH yielded a positive slope of 0.18 and a $p$-value of 0.248, indicating a weak positive correlation. However, the $R^{2}$ value remained low at 0.14, which suggests significant scatter in the data. This weak positive correlation suggests that the chemical link between \ce{CH3OH} and HCOOH might not be possible. This also indicates that the formation mechanism of HCOOH depends not only on the \ce{CH3OH} but also on the other O-bearing molecules and local physical conditions that regulate grain-surface reaction networks and desorption mechanisms.

\begin{table}
	\centering
	\caption{Peak fractional abundances and temperatures of HCOOH, \ce{CH3OH} and \ce{H2CO} predicted by the model under fast warm-up conditions.}
	\begin{adjustbox}{width=0.48\textwidth}
		\begin{tabular}{lcccc}
			\hline
			Species         & Phase          & Temperature (K) & Abundance              \\
			\hline
			\multirow{3}{*}{HCOOH}   
			& Gas     & 127 & $2.92\times10^{-9}$ \\
			& Surface & 102  & $1.26 \times 10^{-10}$  \\
			& Mantle    & 18  & $3.81 \times 10^{-10}$  \\
			\hline
			\multirow{3}{*}{\ce{CH3OH}}     
			& Gas    & 130  & $2.82 \times 10^{-7}$ \\
			& Surface & 110             & $3.27 \times 10^{-8}$  \\
			& Mantle    & 10            & $8.51 \times 10^{-8}$  \\
			\hline
			\multirow{3}{*}{\ce{H2CO}}     
			& Gas    & 86  & $3.21 \times 10^{-9}$\\
			& Surface & 55             & $1.32 \times 10^{-10}$  \\
			& Mantle    & 10            & $6.21 \times 10^{-10}$  \\
			\hline
		\end{tabular}
	\end{adjustbox}
	\label{tab:abundances}
\end{table}

\begin{table*}
	\centering
	\caption{Comparison between peak gas-phase fractional abundances in the warm-up stage and the observed values.}
	\begin{adjustbox}{width=1.0\textwidth}
		\begin{tabular}{|c|cc|cc|c|c}
			\hline 
			Molecule & Modelled abundance & Modelled temperature& Observed abundance & Excitation temperature & \multicolumn{1}{c|}{Dominant reactions} \\
			&                  &  (K)                 &                  &           (K)             &      \\
			\hline
			\ce{CH3OH}/\ce{H2} &2.82$\times$10$^{-7}$ &130 &(1.68$\pm$0.14)$\times$10$^{-7}$ &152$\pm$16 &\ce{CH3O} + H $\rightarrow$ \ce{CH3OH} \\
			\hline
			\ce{H2CO}/\ce{H2}  &3.21$\times$10$^{-9}$  &86 &(2.26$\pm$0.19)$\times$10$^{-9}$&90$\pm$6& \ce{HCO} + H $\rightarrow$ \ce{H2CO} \\
			\hline
			HCOOH/\ce{H2}      &2.92$\times$10$^{-9}$  &127 &(2.62$\pm$0.29)$\times$10$^{-9}$ &120$\pm$15 &\ce{HCO}+OH $\rightarrow$ HCOOH\\
			\hline
			HCOOH/\ce{CH3OH}   &1.03$\times$10$^{-2}$  &-- &(1.56$\pm$0.12)$\times$10$^{-2}$ &-- &-- \\
			\hline
			HCOOH/\ce{H2CO}    &0.91  & --&(1.16$\pm$0.12) &-- &-- \\
			\hline
		\end{tabular}	
	\end{adjustbox}
	\label{tab:abun}
\end{table*}

\subsection{Chemical modelling of HCOOH and other molecules}
We computed a three-phase (gas + grain-surface + icy mantle) warm-up chemical model of HCOOH, \ce{CH3OH}, and \ce{H2CO} using the time-dependent gas-grain chemical code UCLCHEM \citep{vi13, hold17} to investigate the modelled abundance and potential production mechanisms of these molecules. This chemical code investigates thermal and non-thermal desorption in both gas-phase and grain surface chemistry, considering various sources under different physical conditions in the ISM. This chemical code solves the rate of reactions to estimate the fractional abundances relative to hydrogen in the gas-phase and grain surface molecules in different environments where molecules exist \citep{vi13}. In accordance with \citet{gar06}, we employed a two-stage physical model to explain the physical conditions in HMCs: free-fall collapse followed by a dynamically static warm-up. In the cold collapse stage (Phase I), the density of the gas ($n_{H_2}$) increases from $1\times10^{2}$ cm$^{-3}$ to $1\times10^{8}$ cm$^{-3}$, and the gas-grain temperature is fixed at 8 K. In chemical modelling, the cosmic ray ionization rate ($\zeta$) and initial visual extinction ($A_{V}$) are $1.3\times10^{-17}$ s$^{-1}$ and 2. During this period, the accretion rate of atoms and molecules on the grain surface is $10^{-5}~\textup{M}_{\odot}$ yr$^{-1}$, and it depends on the gas density of the hot cores \citep{viti04}.  We assumed the sticking probability during chemical modelling to be unity, meaning that all incoming hydrogen atoms will stick to the grain surface if they encounter an inactive site. The molecular species may hydrogenate or undergo rapid reactions with other species on the grain surface during this period.  In the chemical model, the abundance of atoms relative to the solar values, which were used at the beginning of the collapse phase, is shown in Table~\ref{tab:mod}, which is taken from \citet{gar13}. 
	
In phase II (warm-up stage), the temperature is increased from 8 K to 400 K, where $n_{H_2}$ remains constant at $1\times10^{8}$ cm$^{-3}$. We adopted the fast warm-up model with a timescale of $7.12\times10^{4}$ yr, which was taken from \citet{gar13}. Previous studies show that the fast warm-up model is well-suited to the physical conditions of high-mass star-forming regions \citep{suz18}. In this model, the warm-up phase lasts for $7.12\times10^{4}$ yr, after which the temperature remains constant at its peak value of 400 K. In this stage, the temperature increases via the following expression \citep{gar13}.

\begin{equation}
T = T_0 + (T_{max}-T_0)(\Delta t/t_h)^n
\label{tab:temp}
\end{equation}
In Equation~\ref{tab:temp}, $T_{0}$ = 8 K represents the initial temperature at the start of the warm-up phase, and $T_{max}$ denotes the maximum temperature reached at the end of the warm-up phase. The parameter $\Delta t$ corresponds to the time duration of the warm-up phase, and $t{_h}$ represents its characteristic timescale in the warm-up stage. The warm-up stage follows a first-order process, with $n$ = 1. In this study, we set the fiducial values of $t{_h} = 7.12\times10^{4}$ yr and $T_{max}$ = 400 K. During the warm-up phase, we also incorporated C-type shocks with a shock velocity of $v_s = 20$ km s$^{-1}$, as adopted from \citet{ao12} and \citet{jim18}. The inclusion of C-type shocks is motivated by the observed outflow signatures in both \ce{H2CO} and \ce{CH3OH}. It is believed that the temperatures of the gas and dust are closely associated during the warm-up stage. The molecules no longer freeze during the warm-up phase, and both thermal and non-thermal desorption processes leave frozen molecules on the grain surface in the gas-phase. Furthermore, we incorporated volcanic desorption, monomolecular desorption, grain mantle desorption, and co-desorption with \ce{H2O} into our model (see Tables~1 and 2 of \citet{viti04}). This chemical model is similar to that of \cite{gar13}, \citet{man24a}, \cite{man24d}, and \citet{man24e}.

During chemical modelling, we used more than 300 reactions of HCCOH, \ce{CH3OH}, and \ce{H2CO}, which were taken from \cite{gar06}, \cite{gar08}, \cite{gar13}, and the astrochemical databases of KIDA \citep{wak12} and UMIST \citep{mc13}. In the chemical model, the binding energies (E$_{D}$) of HCOOH, \ce{CH3OH}, and \ce{H2CO} were  6000 K,  5000 K, and  4500 K, respectively. The  E$_{D}$ values of HCOOH, \ce{CH3OH}, and \ce{H2CO} were obtained from \cite{suz18}. The behaviour of the fractional abundances relative to hydrogen of HCCOH, \ce{CH3OH}, and \ce{H2CO} over a time scale of $7.12\times10^{4}$ yr is illustrated in Figure~\ref{fig:model}. Table~\ref{tab:abundances} displays the peak abundances and the temperatures of the molecules. The formation of \ce{H2CO} and \ce{CH3OH} in the gas-phase is minimal; instead, it primarily occurs on dust grains through successive hydrogenation of CO accreted from the gas-phase. The dominant reaction of the formation pathway of \ce{H2CO} and \ce{CH3OH} on the dust grains is,\\\\
CO$\stackrel{H}{\longrightarrow}$HCO$\stackrel{H}{\longrightarrow}$\ce{H2CO}$\stackrel{H}{\longrightarrow}$\ce{CH3O}$\stackrel{H}{\longrightarrow}$\ce{CH3OH}~~~~~~~~(2)\\\\
The high gas-phase abundance of \ce{CH3OH} at low temperatures primarily results from chemical desorption from dust grains. In the simulation, a moderate increase in \ce{CH3OH} abundance occurred around $T\sim$ 40 K because of the electronic recombination of \ce{HC(OH)OCH3}$^{+}$, formed from the reaction between protonated methanol (\ce{CH3OH2}$^{+}$) and \ce{H2CO} when \ce{H2CO} was abundantly released into the gas-phase. During simulation, we observed that the abundance of HCOOH in the gas-phase increased when \ce{H2O} evaporated, thus improving the formation of OH through its reaction with atomic oxygen. The OH radical then reacted with \ce{HCO} in the grain surface to produce HCOOH. The dominant reaction of the formation pathway of HCCOH on the grain surface is \\\\
\ce{HCO} + OH $\longrightarrow$ HCOOH ~~~~~~~~(3)\\\\
At this stage, the grain surface synthesis of reaction 3 remains approximately ten times faster owing to the high \ce{HCO} abundance. The final peak in the grain surface HCOOH abundance, along with the slight increase before the last gas-phase peak, is driven by enhanced gas-phase formation as OH desorbs from the dust grains. Eventually, surface-bound HCOOH evaporates, resulting in a peak gas-phase abundance of $2.92\times10^{-9}$ relative to \ce{H2}. Overall, grain surface reactions between \ce{HCO} and OH are the primary pathway for HCOOH formation.

\subsection{Comparison between observed and modelled abundances of HCOOH}
We compared the observed abundance of HCOOH with the modelled abundance derived from three-phase warm-up chemical modelling using UCLCHEM. This comparison aims to investigate the possible formation pathways of HCOOH towards G358.93--0.03 MM1. Physically, this comparison is suitable because the gas density of G358.93--0.03 MM1 is $1.08\times10^{8}$ cm$^{-3}$, which is similar to the chemical model gas density. The comparison between the observed and modelled abundances is shown in Table~\ref{tab:abun}. We found that the observed abundance of HCOOH relative to \ce{H2} is nearly similar to the modelled value within a factor of 0.89. This indicates that HCOOH might be formed through the reaction between \ce{HCO} and OH (Reaction 3) on the grain surface, which is further released in the gas-phase. We also found that the observed and modelled abundance of \ce{CH3OH} and \ce{H2CO} are nearly close, with the modelled values within the factors of 0.59 and 0.70, respectively. That indicates \ce{H2CO} and \ce{CH3OH} may be formed via the hydrogenation of CO on the grain surface (Reaction 2). The previous chemical model by \citet{gar06} also indicates that reactions 2 and 3 are the dominant pathways for the formation of \ce{CH3OH}, \ce{H2CO}, and HCOOH on the grain surface. The observed and modelled abundance ratios of HCOOH/\ce{CH3OH} and HCOOH/\ce{H2CO} are in good agreement, differing by factors of only 1.51 and 1.27, respectively. Since the observed and modelled abundances, including temperatures, of three molecules are nearly close, this suggests that our chemical model is well-suited for the G358.93--0.03 MM1 hot core.

\section{Summary and conclusion}
\label{sec:con}
In this article, we present the first detection of rotational emission lines of $t$-HCOOH towards hot core G358.93--0.03 MM1 using ALMA band 7. The column density and excitation temperature of $t$-HCOOH are $(8.13\pm0.72)\times10^{15}$ cm$^{-2}$ and 120 $\pm$ 15 K, respectively. The fractional abundance of $t$-HCOOH relative to H$_{2}$ is $(2.62\pm0.29)\times10^{-9}$. The column density ratios of $t$-HCOOH/\ce{CH3OH} and $t$-HCOOH/\ce{H2CO} are $(1.56\pm0.12)\times10^{-2}$ and $(1.16\pm0.12)$, respectively. From the molecular correlation analysis, we find a negative correlation between HCOOH and \ce{H2CO}, while HCOOH and \ce{CH3OH} show a positive but statistically insignificant correlation. This suggests that \ce{H2CO} and \ce{CH3OH} may not serve as effective precursors of HCOOH. To understand the possible formation pathways of HCOOH and other molecules, we computed a three-phase warm-up chemical model using the gas-grain chemical code UCLCHEM. We found that the observed and modelled abundances of HCOOH are almost similar within a factor of 0.89. From these results, we claim that HCOOH might be formed through the reaction between \ce{HCO} and OH on the grain surface, which is further released in the gas-phase. In addition to HCOOH, we also detected several other COMs, which we discuss in follow-up studies.

\begin{acknowledgements}
We thank the anonymous reviewers for their helpful comments, which improved the manuscript. S.S. expresses gratitude to the Swami Vivekananda Merit-cum-Means Scholarship (SVMCM) for providing financial support for this research. This paper makes use of the following ALMA data: ADS/JAO.ALMA\#2019.1.00768.S. ALMA is a partnership of ESO (representing its member states), NSF (USA), and NINS (Japan), together with NRC (Canada), MOST and ASIAA (Taiwan), and KASI (Republic of Korea), in cooperation with the Republic of Chile. The Joint ALMA Observatory is operated by ESO, AUI/NRAO, and NAOJ.
\end{acknowledgements}

\bibliographystyle{raa}

\end{CJK*}
\appendix
\renewcommand{\thefigure}{A\arabic{figure}}  
\setcounter{figure}{0}                       
\renewcommand{\thetable}{A\arabic{table}}  
\setcounter{table}{0}

\section*{Appendix}

\begin{figure*}
	\centering
	\includegraphics[width=1.0\textwidth]{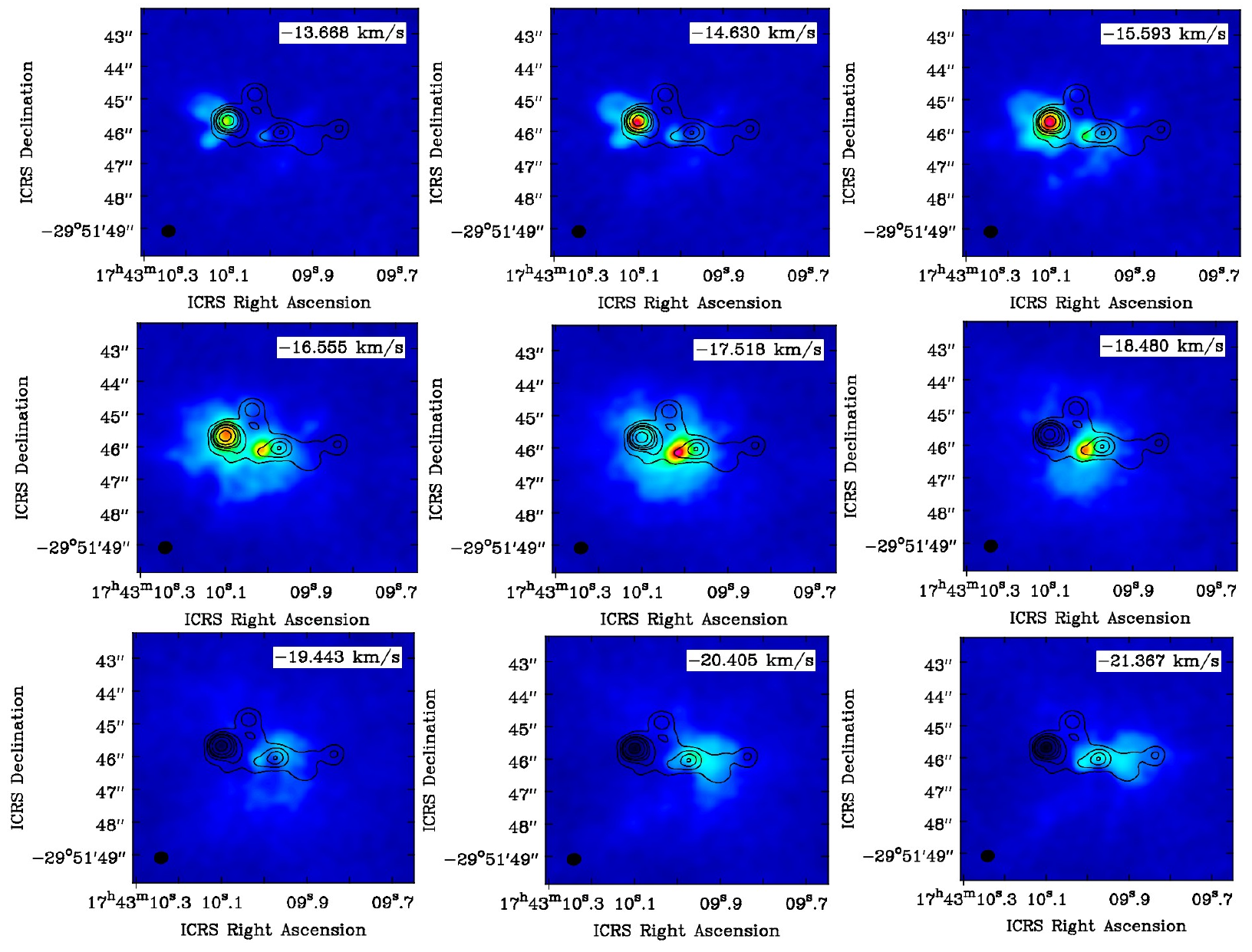}
	\includegraphics[width=0.5\textwidth]{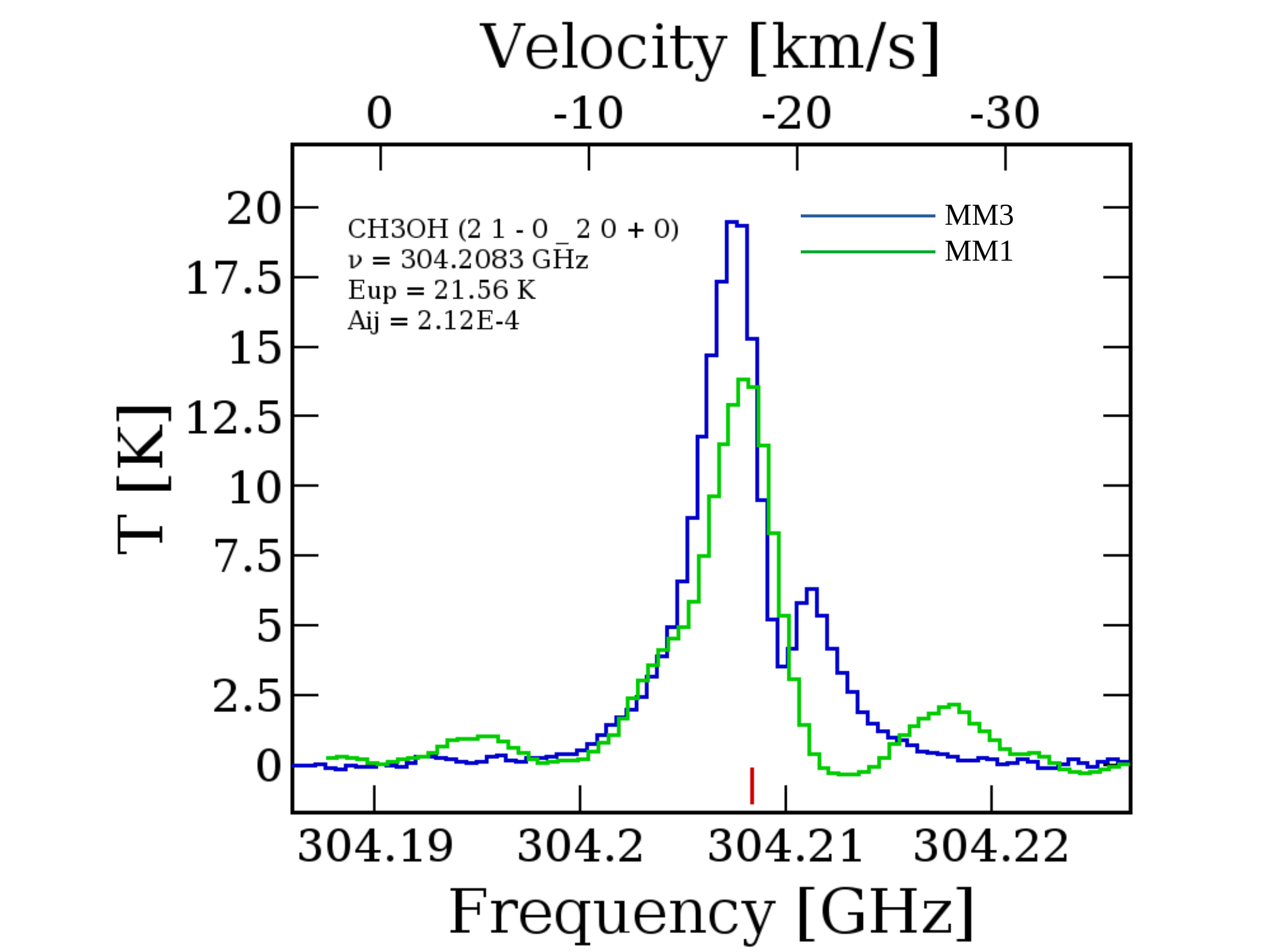}
	\caption{Channel maps of \ce{CH3OH} with the transition $J$ = 2(1,-0)--2(0,+0) toward G358.93--0.03 are shown in the upper panel. The lower panel presents the corresponding emission spectra of \ce{CH3OH} ($J$ = 2(1,-0)--2(0,+0)) toward the G358.93--0.03 MM1 and MM3 cores.}
	\label{fig:CH3OHCHANNEL}
\end{figure*}

\begin{figure*}
	\centering
	\includegraphics[width=1.0\textwidth]{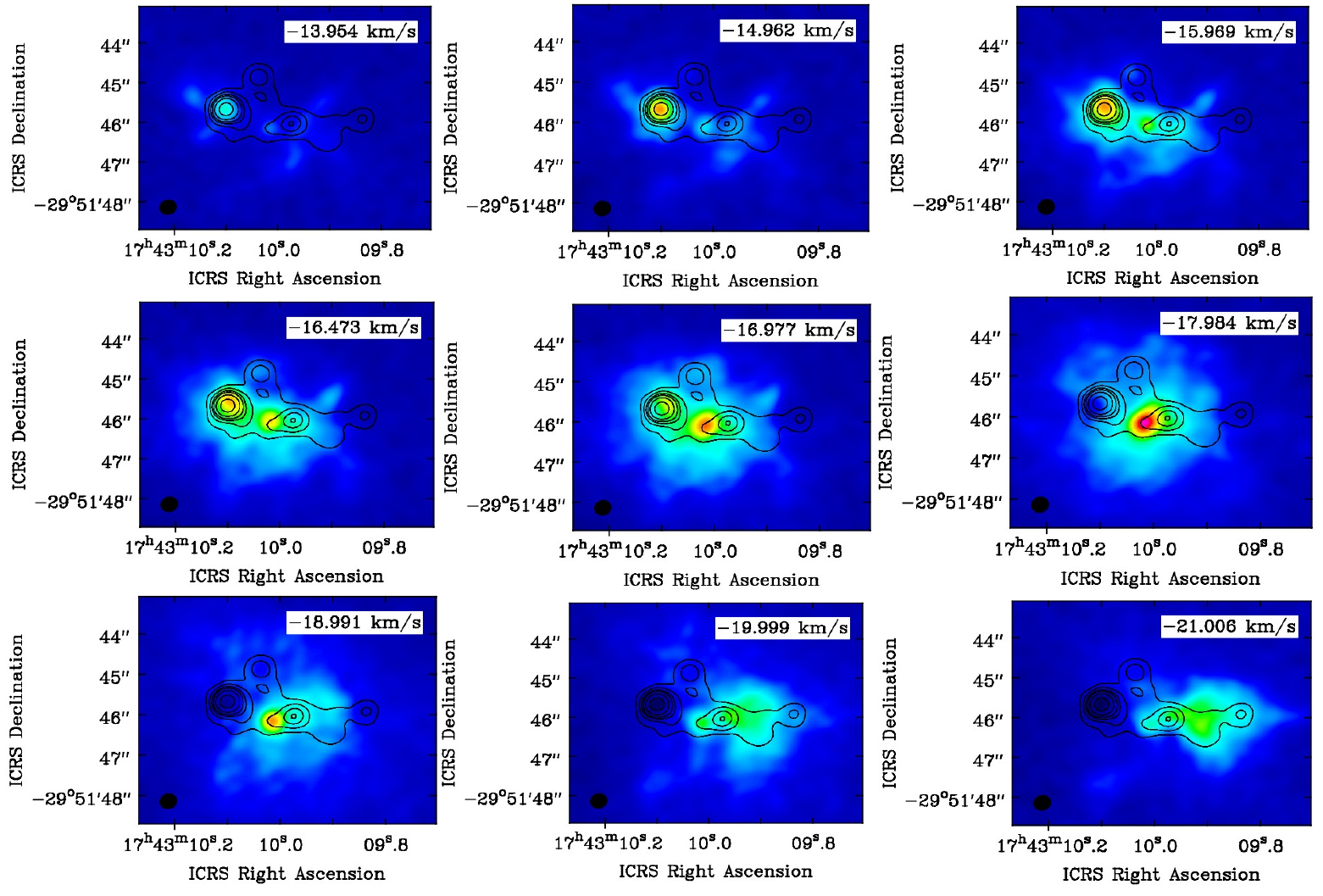}
	\includegraphics[width=0.5\textwidth]{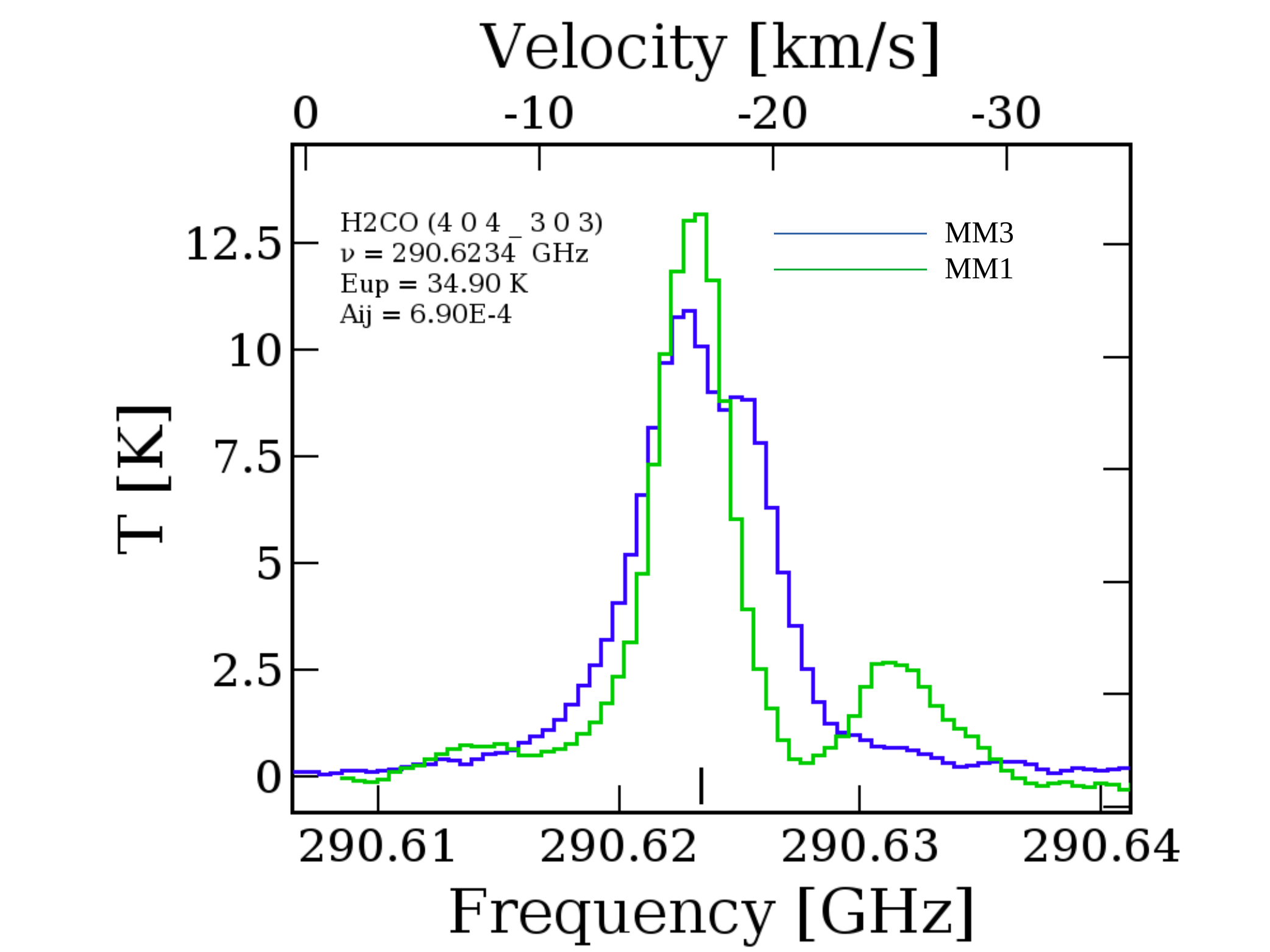}
	\caption{Channel maps of \ce{H2CO} with the transition $J$ = 4(0,4)--3(0,3) toward G358.93--0.03 are shown in the upper panel. The lower panel presents the corresponding emission spectra of \ce{H2CO} ($J$ = 4(0,4)--3(0,3)) toward the G358.93--0.03 MM1 and MM3 cores.}
	\label{fig:H2COCHANNEL}
\end{figure*}

\end{document}